\documentclass[twoside]{article}

\usepackage[T1]{fontenc} 
\usepackage[english]{babel} 
\usepackage[hmarginratio=1:1,top=32mm,columnsep=20pt]{geometry} 
\usepackage{booktabs} 
\usepackage{enumitem} 
\setlist[itemize]{noitemsep} 
\usepackage{abstract} 

\usepackage{titlesec} 
\renewcommand\thesection{\Roman{section}} 
\renewcommand\thesubsection{\roman{subsection}} 
\titleformat{\section}[block]{\large\scshape\centering}{\thesection.}{1em}{} 
\titleformat{\subsection}[block]{\large}{\thesubsection.}{1em}{} 

\usepackage{fancyhdr} 
\usepackage{titling} 

\usepackage{hyperref} 

\usepackage{tikz-cd}
\usepackage[all,cmtip]{xy}
\usepackage{amsmath,amssymb}
\usepackage{graphicx}
\usepackage[english]{babel}
\usepackage[T1]{fontenc}
\usepackage[utf8]{inputenc}

\newcommand{\ac}{\mathrm{ac}}

\newcommand{\ad}{\mathop{\mathrm{Ad}}\nolimits}
\newcommand{\at}{\mathop{\mathrm{At}}\nolimits}

\newcommand{\Body}{\mathcal{B}}
\newcommand{\Curv}{\mathop{\mathrm{Curv}}}

\newcommand{\con}{{\mathrm{CP}}}
\newcommand{\dd}{\mathrm{d}}
\newcommand{\Div}{\mathop{\mathrm{div}}\nolimits}
\newcommand{\EL}{\mathcal{EL}}
\newcommand{\End}{\mathop{\mathrm{End}}\nolimits}
\newcommand{\EP}{\mathcal{EP}}
\newcommand{\ES}{\mathbb{E}^3} 
\newcommand{\fp}[1]{{\mathop{\times}}_{#1}}

\newcommand{\Gau}{\mathop{\mathrm{Gau}}\nolimits}
\newcommand{\Str}[1]{#1\mathrm{Str}}

\newcommand{\Id}{\mathop{\mathrm{Id}}\nolimits}

\newcommand{\Int}{\mathop{\mathrm{Int}}\nolimits}
\newcommand{\Ext}{\mathop{\mathrm{Ext}}\nolimits}
\newcommand{\Fr}{\mathop{\mathrm{Fr}}\nolimits}

\newcommand{\La}{\mathcal{L}}

\newcommand{\Lie}[1]{{\mathop{\mathrm{Lie}}#1}}
\newcommand{\LL}{\mathbb{L}}

\newcommand{\parcial}[2]{\frac{\partial #1}{\partial #2}}

\newcommand{\RR}{\mathbb{R}}
\newcommand{\Star}{\mathrm{Star}}
\newcommand{\Cl}{\mathop{\mathrm{Cl}}\nolimits}
\newcommand{\vol}{\mathrm{vol}}

\newcommand{\ZZ}{\mathbb{Z}}

\newtheorem{remark}{Remark}
\numberwithin{remark}{section}

\numberwithin{example}{section}

\numberwithin{lemma}{section}
\newtheorem{theorem}{Theorem}
\numberwithin{theorem}{section}

\numberwithin{corollary}{section}
\newtheorem{define}{Definition}
\numberwithin{define}{section}

\newtheorem{prop}[theorem]{Proposition}

\pretitle{\begin{center}\Huge\bfseries} 
\posttitle{\end{center}} 
\title{Discrete Formulation for the dynamics of rods deforming in space} 
\author{%
\textsc{Ana Casimiro}\thanks{Supported by Fundação para a Ciência e a Tecnologia (Portuguese Foundation for
	Science and Technology) through the project UID/MAT/00297/2013 (Centro de Matemática e Aplicações)} \\[1ex] 
\normalsize CMA - Centro de Matem\'{a}tica e Aplica\c{c}\~{o}es, Departamento de Matem\'{a}tica,\\
Faculdade de Ci\^{e}ncias e Tecnologia, Universidade Nova de Lisboa,\\ Quinta da Torre 2829-516 Caparica, Portugal \\ 
\normalsize {amc@fct.unl.pt} 
\and 
\textsc{César Rodrigo}\thanks{Supported by Fundação para a Ciência e a Tecnologia (Portuguese Foundation
	for Science and Technology) through the project UID/MAT/04561/2013 (Centro de Matemática, Aplicações Fundamentais e Investigação Operacional of
	Universidade de Lisboa CMAF-CIO).} \\[1ex] 
\normalsize CMAF-CIO - Centro de Matem\'{a}tica, Aplica\c{c}\~{o}es Fundamentais e Investiga\c{c}\~{a}o Operacional,\\ CINAMIL, Academia Militar,\\ Av.~Conde Castro Guimar\~{a}es, 2720-113 Amadora, Portugal \\ 
\normalsize {crodrigo@geomat-pt.com} 
}
\date{Sent for Publication in Journal of Mathematical Physics on June 16th 2018} 
\renewcommand{\maketitlehookd}{%
\begin{abstract}
\noindent The movement of rods in an euclidean space can be described as a field theory on a principal bundle. The dynamics of a rod is governed by partial differential equations that may have a variational origin. If the corresponding smooth lagrangian density is invariant by some group of transformations, there exist corresponding conserved Noether currents. Generally numerical schemes dealing with PDEs fail to reflect these conservation properties.

We describe the main ingredients needed to create, from the smooth lagrangian density, a variational principle for discrete motions of a discrete rod, with corresponding conserved Noether currents. We describe all geometrical objects in terms of elements on the linear Atiyah bundle, using a reduced forward difference operator. We show how this introduces a discrete lagrangian density that models the discrete dynamics of a discrete rod. The presented tools are general enough to represent a discretization of any variational theory in principal bundles, and its simplicity allows to perform an iterative integration algorithm to compute the discrete rod evolution in time, starting from any predefined configurations of all discrete rod elements at initial times.

\em{Keywords: Variational Integrator, Reduction, Discretization, Euler-Poincaré, Elastic Rods}
\end{abstract}
}

\begin{document}

\maketitle


	\section{Introduction}
In the present work we shall follow the specific case of the dynamical equations of hyperelastic rods as leitmotiv to report certain results concerning the discretization of variational principles in principal bundles. These results were described by the authors in \cite{CasiRodr17}  and \cite{CasiRodr17b}.

Motivated by the large spreading of geometrical techniques that generate variational integrators with good conservation properties in mechanical systems \cite{BobeSuri99,CasiRodr12,CasiRodr12b,ChriMuntOwre11,CortMart01,DemGayLeyObeRatWei,DemoGayRat,FerGarRod12,GawlMullPavlMarsDesb11,GuoWu03,HaiLubWan06,KoMa10,Kobilarov14,McLaQuis06,LeonDiegSant04,LewMarsOrtiWest03,MarsWest01,Vanker07,VankCantr07,WendMars97}, the question arises up to which point one may construct a category of arbitrary discrete variational principles, with similar geometrical structures to existing smooth counterparts, broad enough to include field theories and mechanics, with or without nonholonomic constraints, and to admit geometrical considerations such as the notion of symmetry or the notion of reduction by some group action. The prize for such a work is a discrete formalism, paralleling the existing one of smooth variational principles, that are so common in physical theories. As companion work to this construction of a discrete variational formalism, one should answer certain questions concerning the elements contained in it. First of all if these elements may be related to other objects that arise in the smooth variational formalisms, and what tools are needed to establish such relations, in a nonlinear situation. 	

As a result we present here a successful approach to the discretization of a variational theory that involves a field theory with several dependent and independent variables on arbitrary manifolds, in a situation that includes non-trivial bundles with non-commutative groups of symmetries, where a reduction procedure allows to describe critical fields through a system of partial differential equations formulated for a reduced field. More precisely, we shall deal with a general problem of Euler-Poincaré reduction for a field theory in a principal bundle \cite{CasChaGar13,CasGarRat01,CasGarRod13}, a common situation in several scientific branches, like motion and control of mechanical systems \cite{KoMa10,Kobilarov14}, elasticity\cite{DemGayLeyObeRatWei,DemoGayRat}, liquid crystal theory \cite{GayRatTro}, Palatini gravity \cite{Capr14}, and other gauge field theories. We  show how to formulate the discrete counterparts of the smooth elements present in the theory and in its reduction, and present the needed objects that allow to relate smooth and discrete fields. The proposed formalism differs from previous ones \cite{Castrillon12,FernZucc13,Leok05,LewMarsOrtiWest03} for its geometrical origin, independence of trivialization choice and validity for any nontrivial principal bundle. Finally we shall describe an algorithm that integrates the discrete partial difference equations that arise in the discrete reduced formalism.

We shall assume on the reader a familiarity with the main geometrical tools of fiber bundle theory and shall present elastic rod theories within this language. As a guiding reference for the foundations of the geometrical theory of elastic bodies moving on space we shall follow the reference \cite{Antman94}, in particular chapter 8 (``Theory of rods deforming in space''). We denote by $\ES$ the 3-dimensional euclidean space and $d_E$ the corresponding euclidean distance function. We shall consider a body $\Body$ to be a set, whose elements $\alpha\in\Body$ we call material points or atoms of the body. There exists a measure $\mu$ on $\mathcal{B}$, determining the total mass $\mu(D)$ of measurable domains $D\subset\Body$.  A configuration of the body in $\ES$ (or briefly a configuration) is any specific immersion $p\colon \Body\hookrightarrow \ES$.  It is convenient to fix some (admissible) reference configuration $\widehat{p}\colon \Body\hookrightarrow \ES$ so that $\Body$ may be identified as a certain subset of points of the euclidean space. In many cases a topological, differential or metric structure is fixed in the body and a certain subset of admissible immersions (that includes $\widehat{p}$) is chosen, demanding that the immersion is a continuous, smooth or distance-preserving mapping. Dynamics deals with time-evolution of such configurations, hence with trajectories $p(t)$ determining some (admissible) configuration for each time instant $t\in T$ in a certain continuous time interval $T\subset \RR$.

In the case of bodies with a finite number $N$ of atoms, the set of configurations can be identified with $(\ES)^N$, a smooth manifold with dimension $3N$, and admissible configurations with some submanifold. For such bodies a dynamical theory can be given with the usual tools of geometrical mechanics (with holonomic or non-holonomic constraints) on this large-dimensional manifold. Discretization techniques available in geometrical mechanics can be used: a choice of a finite monotone sequence of time events is given $J_t=(t_j)_{j\in J}\subset\RR$, indexed by some monotone mapping $j\mapsto t_j$ defined on elements $j\in J$ of some interval $J\subset\ZZ$ of integer values. There arises a corresponding theory on discrete trajectories $(p_t(j))_{j\in J}$, interpreted as approximations of $p(t_j)$ for some smooth trajectory $p(t)$ and monotone sequence $(t_j)_{j\in J}$. The discrete theory can be used to formulate algorithms that use initial data to determine specific discrete trajectories $(p_t(j))_{j\in J}$, and to measure its error when considered as approximate values of $p(t_j)$, for the smooth evolution $p(t)$ of the body. However this specific mechanical theory is large-dimensional and, in the case of rods, does not take advantage of the particular filament-like structure or the possibility to work with reduced coordinates.

As mentioned, a rod is a particular kind of body with a filament-like structure: it admits a partition into slices $\Body=\sqcup_{s\in S}\Body_s$ where $s\in S$ is a real parameter ranging in a certain bounded subset $S\subset \RR$. Such structure can be given, for example, using the reference configuration and a particular function $f$ on $\ES$ (smooth one, or just with a discrete set of values, in the discrete case). The slice $\Body_s$ associated to a paramenter $s$ is given by atoms whose reference configuration lies on the level set $f=s$. In the case of rods with a finite number of atoms we may assume that the rod structure is given by some finite set of slices that can be ordered in a finite monotone sequence $I_s=\{s_1<s_2<\ldots <s_m\}\subset \RR$, indexed by a mapping $i\in I\mapsto s_i\in S$ for some bounded interval $I\subseteq \ZZ$ of integer values, and that each slice is a nonempty finite set with $n_i=\sharp \Body_{s_i}$ atoms (a monomer with $n_i$ atoms, if the body is seen as a polymer). In this case the body has $N=\sum_{i\in I} n_i$ atoms, and configurations form a manifold with dimension $3N$.

In rod theory only a certain (nonempty) subset of configurations $p\colon \mathcal{B}\rightarrow \ES$ shall be considered as admissible. Apart from smoothness of the configuration (when the body has a smooth structure), a reasonable specific way to restrict our family of configurations arises when each slice is assumed to be rigid, that is, each slice $\Body_s$ is endowed with a distance function $d_s\colon \Body_s\times\Body_s\rightarrow \RR$, and the only admissible configurations are $p\colon\Body\hookrightarrow \ES$ such that $d_E(p(\alpha),p(\tilde\alpha))=d_s(\alpha,\tilde \alpha)$ for each pair of atoms sharing the same slice $\alpha,\tilde \alpha\in\Body_s$. In this way we fix a subset of configurations. Admissible configurations $p$ can be seen as sequences $\left(p_s\right)_{s\in S}$ of isometric immersions $p_s\colon\Body_s\hookrightarrow \ES$, for each slice, into euclidean space. Such mappings $p\colon s\mapsto p_s$ can be defined geometrically as sections of a bundle $\pi\colon P\rightarrow S$ (that is, $\pi\circ p=\Id_S$):
\begin{define}
	For a fixed rod $(\Body_s,d_s)_{s\in S}$, we call bundle of admissible rod configurations the bundle $\pi\colon P\rightarrow S$, where the fiber at $s\in S$ is the manifold $P_s$ of isometric immersions  $p_s\in\mathrm{Isom}(\Body_s,\ES)$. The space of such sections is denoted by $\Gamma(P)$. For a given interval $T\subset\RR$ we call bundle of admissible rod motions in time $T$ the bundle $P\times T\rightarrow S\times T$. Each fiber $P_s$ (or $P_{s,t}$) is called the space of admissible configurations of a slice $\Body_s$ (at time $t$).
\end{define}

For the particular case of homogeneous rods we may assume that all slices $\Body_s$ to be identical, that is, there exists a body $\mathcal{S}$ (the model slice) with metric structure, and a canonical isometrical identification $\mathcal{S}\simeq \Body_s$, so that $\Body=\mathcal{S}\times I$, $P_s=\mathrm{Isom}(\mathcal{S},\ES)$. In this case the bundle of configurations is $P=I\times \mathrm{Isom}(\mathcal{S},\ES)$, a trivial bundle with fiber $\mathrm{Isom}(\mathcal{S},\ES)$.

Motions of any rod are determined then as trajectories $t\in T\mapsto p(t)\in \Gamma(P)$, or equivalently, as sections $(s,t)\mapsto (p_s(t))_{s\in S}$ of the bundle $P\times T\rightarrow S\times T$.

There exists a left action $\lambda\colon G\times P_s\rightarrow P_s$ of the euclidean group of rigid motions $G=\mathrm{Isom}(\ES,\ES)$ on each fiber of the bundle of admissible rod configurations or motions, by composition $\lambda(g,p_s)= g\circ p_s$, and consequently an action on the sections of these bundles: $p_s(t)\mapsto g\circ p_s(t)$

If we fix an euclidean referential on $\ES$, we may represent its points by corresponding homogeneous coordinates $r^i\colon \ES\rightarrow \RR$, normalized so that $r^4=1$, in this case  each point $p\in\ES$ can be represented by a matrix $(r^i)_{i=1\ldots 4}\in M_{4\times 1}(\RR)$ with $r^4=1$. Also using the referential, any euclidean motion $g\in G$ is determined by a matrix $C=\left[\begin{matrix} A&b\\ 0&1\end{matrix}\right]$ where $A\in SO(3)$, $b\in M_{3\times 1}(\RR)$ and the action is matrix product: $\lambda(g,p)\in \ES$ has normalized homogeneous coordinates $\tilde r^i$ given by  $\tilde r=C\cdot r\in M_{4\times 1}(\RR)$.

A slice configuration $p_s\in P_s$ is determined by a matrix $r_s=(r^i_\alpha)_{\alpha\in \Body_s}^{i=1,2,3,4}\in M_{4\times n_i}(\RR)$ with all values 1 on the lower row. If $g\in G$ is represented by a matrix $C$, then $\lambda(g,p)$ is represented by the matrix product $C\cdot r_s$

We assume all slices to be oriented 3-dimensional bodies, that is, there really exists an isometric immersion $p_s\colon\Body_s\hookrightarrow \ES$ and the isotropy group of a configuration $p_s$ is trivial (we mean there is no isometry on the euclidean space that acts as identity on $p_s(\Body_s)$, unless it is the identity $\Id\colon \ES\rightarrow \ES$). In this case the action of the group $G$ on the bundle $P$ is free, proper action and the bundle projector $\pi\colon P\rightarrow S$ is identified with the quotient mapping $\pi^G\colon P\rightarrow P/G$. The bundle of admissible rod configurations is then a principal $G$-bundle. The same holds for the bundle of admissible rod motions $\pi\colon P\times T\rightarrow S\times T$. Dynamics of rods deforming in space is then a particular case of field theories on a principal $G$-bundle.

\section{Variational Field theories on principal G-bundles}

We begin with the foundations of jet and connection bundles \cite{Saunders}. For any smooth bundle $\pi\colon Y\rightarrow X$ the linear morphism $\dd\pi\colon TY\rightarrow \pi^*TX$ is surjective and its null-space determines a sub-bundle $\ker\dd\pi=VY\subseteq TY$ and an exact sequence of morphisms of vector bundles on $Y$:
\begin{equation}\label{connectionsequence}
0\rightarrow VY\hookrightarrow TY\rightarrow \pi^*TX\rightarrow 0
\end{equation}
Splittings of this sequence (vector bundle morphisms $J\colon \pi^*TX\rightarrow TY$ such that $\dd\pi\circ J=\Id_{TX}$) are called connections on the bundle $\pi\colon Y\rightarrow X$. Choosing such a splitting is the same as choosing a subspace $H_y\subset T_yY$ with $H_y\oplus V_yY=T_yY$, at each point $y\in Y$.

Linear morphisms $\pi^*TX\rightarrow TY$ are sections of the vector bundle $\pi^*T^*X\otimes TY\rightarrow Y$. The condition of being a splitting is linear and splittings are sections of an affine sub-bundle $JY\subset \pi^*T^*X\otimes TY$, modeled on the vector bundle $\pi^*T^*X\otimes VY$. The bundle $\pi_J\colon JY\rightarrow Y$ is called the bundle of connections associated to $\pi\colon Y\rightarrow X$.

For any fibered morphism $f\colon Y\rightarrow Z$ between two bundles on $X$, the mapping $1\otimes \dd f\colon \pi^*T^*X\otimes TY\rightarrow \pi^*T^*X\otimes TZ$ transforms elements in $JY$ into elements in $JZ$ and defines a mapping $jf\colon JY\rightarrow JZ$. In particular $\pi\colon Y\rightarrow X$ determines $j\pi\colon JY\rightarrow JX\simeq X$, called the first jet bundle associated to $\pi\colon Y\rightarrow X$. Locally defined sections $y\colon X\rightarrow Y$ of $\pi$ induce sections $jy\colon X\rightarrow JY$ of $j\pi$, called jet extensions of $y$. Each element $j_xy\in (JY)_x$ is always the value at $x$ of some jet extension $jy$ for some locally defined section $y\colon X\rightarrow Y$, and any pair $y,\tilde y$ determine the same element $j_xy$ if and only its values and its differentials coincide at the point $x\in X$. The bundle $JY$ is then the natural space to define first order partial differential equations or to give first order lagrangian functions, which is the central element to state some variational principle.

In variational formalism of field theories on a bundle $\pi\colon Y\rightarrow X$ on some $n$-dimensional manifold $X$ \cite{GoldStern73}, lagrangian densities are horizontal $n$-forms on the jet bundle, that is, differential $n$-forms on $JY$ that vanish when applied to any tangent vector vertical for the projection $j\pi\colon JY\rightarrow X$. Such densities can be written as $\La\cdot \vol_X$, for some choice of volume form $\vol_X\in \Omega^n(X)$ on $X$. The function $\La\colon JY\rightarrow \RR$ is called the Lagrangian. For any compact domain $K\subset X$ the lagrangian density $\La\vol_X$ determines an action functional $\LL_K$, defined on sections $y\in\Gamma(K,Y)$ with domain $K$:
\begin{equation}\label{acfun}
\LL_K(y)=\int_K (\La\circ j y)\vol_X
\end{equation}
If $(y_\epsilon)_{0\leq\epsilon\leq \epsilon_{max}}\subset \Gamma(K,Y)$ is a smooth variation of a section $y_0\in\Gamma(K,Y)$ defined on the domain $K$, and using the chain rule we have the expression for the first derivative of $\LL_K(y_\epsilon)$:
\begin{equation*}
\left(\frac{\dd}{\dd\epsilon}\right)_0\LL_K(y_\epsilon)=\int_K \langle \dd^v_{y_0}\La, j\delta_{y_0}\rangle \vol_X,\qquad \delta_{y_0}=\left(\frac{\dd}{\dd\epsilon}\right)_0 y_\epsilon\in\Gamma(y_0^*VY)
\end{equation*}
where $\dd\La\in\Gamma(T^*(JY))$ determines $\dd^v_{y_0}\La\in\Gamma((jy_0)^*V^*(JY))$ using the restriction to the section $j y_0$ and the restriction to $V(JY)\subset T(JY)$, where $\langle\cdot,\cdot \rangle$ represents the duality product with elements in $\Gamma((jy_0)^*V(JY))$, and where $\delta_{y_0}\in\Gamma((y_0)^*VY)\mapsto j\delta_{y_0}\in\Gamma((jy_0)^*V(JY))$ represents the jet extension of vector fields induced by the jet extension $y\in\Gamma(Y)\mapsto jy\in\Gamma(JY)$ of sections.
\begin{define}
	For any local section $y\colon U\rightarrow Y$ of the bundle $\pi\colon Y\rightarrow X$, we call differential at $y$ of the action functional $\LL$, associated to a lagrangian density $\La\vol_X$, the linear functional:
	\begin{equation}\label{dacfun}
	\dd_y\LL\colon \delta_y\in\Gamma^c(y^*VY)\mapsto \int_U\langle \dd^v_{y_0}\La, j\delta_{y}\rangle \vol_X\in\RR
	\end{equation}
	where the subspace $\Gamma^c(y^*VY)$ is given by all sections of the bundle $y^*VY\rightarrow U$ whose support is a compact subset $K\subset U$. 
\end{define}

We may formulate different variational principles with differential constraints on a fibred manifold $\pi\colon Y\rightarrow X$ (see \cite{FernGarcRodr04}). For the case of the fixed boundary variational principle,	an admissible local field $y\in\Gamma(Y)$ is called critical for the variational problem if $\dd_y\LL$ vanishes on the space of admissible infinitesimal variations $\Gamma^c(y^*VY)$. Criticality is then equivalent to the annihilation of the Euler-Lagrange tensor $\EL(y)\in\Gamma(y^*V^*Y)$, that for the case $\vol_X=\dd x^1\wedge\ldots\wedge \dd x^n$ takes the classical form:
\begin{equation}\label{ELeq}
\EL(y)=\left(\parcial{\La}{y^i}(jy)-\frac{\dd}{\dd x^\nu} \left(\parcial{\La}{y^i_\nu}(jy)\right)\right)\dd y^i\in\Gamma(y^*V^*Y)
\end{equation}
the appearance of $(\dd/\dd x^\nu)$ in an object depending on $jy$ shows that equations $\EL(y)=0$ represent a system of second order partial differential equations on the unknown $y\in\Gamma(Y)$.

For a different choice of admissible infinitesimal variations, however, critical sections can be characterized by a different system of partial differential equations. This is, for example, the case of the Euler-Poincaré equations in the case of gauge field theories invariant by the action of a certain group \cite{Vanker07,ElliGayHolmRati11,CasGarRod13}.

\subsection{Ehresmann's gauge groupoid}
Consider a principal $G$-bundle $\pi\colon P\rightarrow X$ with left-action $\lambda\colon G\times P\rightarrow P$ (also denoted by $\lambda_g(p)=\lambda(g,p)=gp$). We may identify elements $x\in X$ with $\pi$-fibers $P_x=\pi^{-1}(x)$ or with $G$-orbits $G\cdot p\subset P$.
This action induces corresponding actions $(\lambda\times \Id)$ on $\pi^*TX$, $\dd\lambda$ on $TP$, and its restriction to $VP$. Splittings for (\ref{connectionsequence}) that covariate with these actions are called principal connections on the principal bundle $\pi\colon P\rightarrow X$. Hence principal connections can be seen as splittings of the following exact sequence of vector bundle morphisms:
\begin{equation*}
0\rightarrow VP/G\hookrightarrow TP/G\rightarrow TX\rightarrow 0
\end{equation*}
The bundle $\pi_{\at}=\at P=TP/G\rightarrow X$ is called Atiyah's vector bundle \cite{Atiyah} associated to $P$. Its sections can be identified with vector fields on $P$ that are invariant for the action $\lambda$ on vector fields of $P$. The bundle $\pi_{\ad}=\ad P=VP/G\rightarrow X$ is a vector sub-bundle of Atiyah's bundle, the null-space of $\dd\pi$, called adjoint vector bundle, its sections are vertical $\lambda$-invariant vector fields on $P$. This exact sequence is called Atiyah's sequence. Its splittings can be seen as sections $\chi$ of $T^*X\otimes \at P$ such that $\dd\pi\circ\chi=\Id_{TX}$. Thus these splittings are sections of the affine bundle $\con\subset T^*X\otimes\at P$ described by linear equations $\dd\pi\circ\chi=\Id_{TX}$. This affine bundle $\pi_{\con}\colon \con \rightarrow P$ is modeled over the vector bundle $T^*X\otimes VP$.

We must observe that the projection to $P$ and the quotient mapping $\pi^G$ determine vector bundle isomorphisms $VP=P\times_X\ad P$, $TP=P\times_X\at P$, $\pi^*TX=P\times_X TX$. Moreover the action $\lambda$ on each bundle $VP$, $TP$, $\pi^*TX$ can be seen as the action $\lambda\times 1$ on each of these fibered products. Any splitting $\chi$ of Atiyah's sequence determines $\Id_P\times \chi$, a splitting of (\ref{connectionsequence}), and this splitting covariates with the actions induced by $\lambda$. Hence any splitting $\chi$ of Atiyah's sequence determines a connection $J$ on $\pi\colon P\rightarrow X$ that covariates with $\lambda$, what we call a principal connection. Conversely, connections $J$ that covariate with $\lambda$ factor by the quotient mappings $\pi^G$ and determine a splitting $\chi$ of Atiyah's sequence. The affine bundle $\pi_{\con}\colon \con\rightarrow P$ is called the bundle of principal connections associated to the principal $G$-bundle $P$.

In the discretization of field theories a natural object used as discrete counterpart of the tangent bundle $\pi_{TY}\colon TY\rightarrow Y$ is the pair bundle $\pi_0\colon (y_0,y_1)\in Y\times Y\mapsto y_0\in Y$. In fact, the $\pi_0$-vertical space of this bundle at a point $(y,y)\in Y\times Y$ has a natural identification $V^{\pi_0}_{(y,y)}(Y\times Y)\simeq T_yY$. The discretization may then use a difference operator, a bundle morphism $\Delta\colon Y\times Y\rightarrow TY$ that generates a difference $\Delta(y_0,y_1)\in T_{y_0}Y$ of two points in a nonlinear manifold \cite{AMS08,Adler,McLa}. This linear object is then more appropriate for the application of classical numerical methods. When we work with principal $G$-bundles and reduce by the action, the natural gauge-independent tangent space is Atiyah's bundle $\at P=TP/G$, and the natural gauge-independent jet space is the bundle of principal connections $JP/G=\con\subset T^*X\otimes \at P$. If $P\times P$ is used as discrete counterpart of $JP$ we would also like to consider $(P\times P)/G$ as discrete counterpart of $\con$. 

Being for any $x\in X$ the action on the fiber $P_x$  free and transitive, for any pair of elements $p_x,\tilde p_x\in P_x$ on the same fiber there exists a unique group element $g\in G$ such that $gp_x=\tilde p_x$. We call this group element the group difference $\tilde p_x p_x^{-1}$ of both configurations. This determines on the fibered product $P\fp{X}P$ a group difference mapping $(p_x,\tilde p_x)\in P_x\times P_x\mapsto \tilde p_x p_x^{-1}\in G$. This mapping however doesn't determine the difference of two arbitrary configurations $(p,\tilde p)\in P\times P$ when they do not share the fiber.

For any specific fiber $x\in X$ we may consider the set of $G$-covariant mappings $\psi\colon P_x\rightarrow P$ whose domain is the single fiber $P_x\subset P$ associated to $x$. Considering all possible fibers, we have a new bundle $\bar\pi_0\colon \End P\rightarrow X$, whose fiber at $x_0$ is:
\begin{equation*}
(\End P)_{x_0}=\{\psi\colon P_{x_0}\rightarrow P\,\text{ domain a single fiber, }\psi\circ\lambda=\lambda\circ(\Id_G\times\psi)\}
\end{equation*}
Giving a section $\psi\colon X\rightarrow \End P$ of this bundle is the same as giving a bundle morphism $\psi\colon P\rightarrow P$ that is $G$-covariant, that is, an endomorphism of $G$-bundles from $P$ to itself, fibered over some mapping $X\rightarrow X$, not necessarily the identity.

As the fibers on $P$ coincide with the $G$-orbits, the image of any such fiber-to-fiber endomorphism is a whole single fiber and there exists  a target mapping $\bar\pi_1\colon \psi\in(\End P)_{x_0}\mapsto \bar\pi_1(\psi)=x_1\in X$ defined by $P_{x_1}=\mathop{\mathrm{Img}}\psi$. We get a pair of projectors $\bar\pi_0\colon \psi\mapsto x_0$ and $\bar\pi_1\colon \psi\mapsto x_1$ (source and target mappings on $\End P$). It is clear that we may compose any pair of elements $\psi,\tilde\psi$ if $\bar\pi_0(\psi)=\bar\pi_1(\tilde\psi)$, and the composition $\psi\circ \tilde\psi$ has source $\bar\pi_0(\tilde\psi)$ and target $\bar\pi_1(\psi)$. The set $\End P$ carries then a groupoid structure, with composition as product and $(\bar\pi_0,\bar\pi_1)\colon \End P\rightarrow X\times X$ as anchor mapping (see \cite{CannWein99} for introductory material on groupoids). This groupoid is also know as Gauge groupoid. However to avoid confusion with the group of gauge transformations, we prefer to denote it as Ehresmann's bundle \cite{Ehresmann50}.

The groupoid $(\bar\pi_0,\bar\pi_1)\colon \End P\rightarrow X\times X$ has a natural groupoid action on the bundle $\pi\colon P\rightarrow X$ by $\ac(\psi,p)=\psi(p)$ whenever $\pi(p)=\bar\pi_0(\psi)$. For notational simplicity, the groupoid action $\ac_\psi\colon p\mapsto \psi(p)$ shall be denoted by $p\mapsto p\psi$, so that commutation with $\lambda_g$ may be stated as $(gp)\psi=g(p\psi)$. We denote then the composition $\psi\circ\tilde\psi$ as $\tilde\psi\psi$. With this notation:
\begin{equation*}
p(\psi\tilde\psi)=(p\psi)\tilde\psi\qquad p\psi=\tilde p\Leftrightarrow p=\tilde p\psi^{-1}
\end{equation*}
Also the groupoid $\End P$ has the induced groupoid action on $VP$ (but not on $TP$!), which comutes with the natural action of $G$, determining so a natural action $\psi\mapsto \ad_\psi$ of the groupoid $\End P$ by linear automorphisms on the vector bundle $\pi_{\ad}\colon \ad P=(VP)/G\rightarrow X$

There exists a natural projection $P\times P\rightarrow \End P$ (the gauge difference mapping) transforming a pair $(p,\tilde p)$ into the unique $G$-covariant transformation $\psi$ defined on the $G$-orbit $Gp$ and such that $\psi(p)=\tilde p$. Namely, $\psi(p_0)=(p_0p^{-1})\tilde p$ for each $p_0\in Gp$. Taking into account the notation $\psi(p_0)=p_0\psi$, we denote this transformation by $\psi=p^{-1}\tilde p$. For this projection the fiber of a single element $\psi\in\End P$ is a $G$-orbit on $P\times P$ by the diagonal action $(\lambda_g\times\lambda_g)(p,\tilde p)=(gp,g\tilde p)$. Hence $(\bar\pi_0,\bar\pi_1)\colon \End P\rightarrow X\times X$ can be naturally identified as the quotient space $(P\times P)/G$ with the mapping induced on the quotient from $(\pi_0,\pi_1)=\pi\times \pi\colon P\times P\rightarrow X\times X$. The gauge difference mapping $(p,\tilde p)\in P\times P\mapsto p^{-1}\tilde p\in\End P$ is the quotient morphism $\pi^G\colon P\times P\rightarrow (P\times P)/G$ expressed in terms of this identification.

The restriction of Ehresmann's bundle to the diagonal $d^1=(\Id,\Id)\colon P\hookrightarrow P\times P$ is the Gauge group bundle $\pi_{\Gau}\colon \Gau P\rightarrow X$ (defined as $\Gau P=(d^1)^*\End P$, that contains only elements $\psi\in\End P$ such that $\bar\pi_0(\psi)=\bar\pi_1(\psi)$). This is a group bundle (each of its fibers is a Lie group) and contains the identity section $Id\colon P\rightarrow \Gau P$. It can be defined also as the quotient bundle $\bar \pi\colon (P\times_XP)/G\rightarrow X$ for the bundle $P\times_XP=(\pi\times\pi)^{-1}(d^1_X)\rightarrow X$, considering the diagonal $d^1_X\subset X\times X$. Moverover, the Lie algebra of each fiber $\Gau P_x$ can be identified with the Lie algebra of $G$-invariant vertical vector fields, which is the fiber of the adjoint bundle $\ad P_x$.

In the case of the bundle of discrete rod configurations $P\rightarrow S$ with discrete set of slices $S=I_s$, for two consecutive slice configurations $p_0,p_1$  (at fibers $s_0<s_1\in S$), the element $p_0^{-1}p_1\in \End P$ can be seen as a determination of the relative position of the configurations of a pair of slices, where by relative position we mean that it remains the same when one applies the same euclidean motion on both slices at the same time. In the case that a reference configuration $\widehat p$ is fixed on each slice, such a relative position $\psi$ can be seen as $\psi=\hat p_0^{-1}g\hat p_1$ for some $g=(\widehat p_0\psi)\widehat p_1^{-1}\in G$, however this euclidean space motion would depend on the particular reference choice, it represents a rigid motion on $\ES$ transforming an orthonormal frame attached to the slice configuration $\widehat p_0(\Body_{s_0})\subset\ES$ into another one attached to the slice configuration $\widehat p_1(\Body_{s_1})\subset\ES$. The element $\psi\in \End P$ should not be mistaken with an euclidean motion $g\in G$, because this representation depends on the reference configuration choice.

Also $\pi^G\colon P\times P\rightarrow (P\times P)/G=\End P$ determines the commutative diagram of exact sequences of vector bundles on $\End P$:
\begin{equation}
\xymatrix{
	0\ar[r] & \mathrm{Sym} \ar@{^{(}->}[r] \ar[d]^{0} &
	(\bar\pi_0,\bar\pi_1)^*(\at P\oplus\at P) \ar[d]^{\overline{\dd\pi}\oplus \overline{\dd\pi}}\ar[r]^{\overline{\dd\pi^G}} & T\End P \ar[d]^{\dd(\bar\pi_0,\bar\pi_1)} \ar[r]& 0\\
	0\ar[r] & 0 \ar@{^{(}->}[r] &(\bar\pi_0,\bar\pi_1)^*(TX\oplus TX) \ar[r]^{\Id} & (\bar\pi_0,\bar\pi_1)^*(TX\oplus TX)\ar[r] & 0
}
\end{equation}
where for each $\psi\in \End P_{x_0,x_1}$ the kernel is $\mathrm{Sym}_\psi=\{(A_0,A_1)\in \ad P_{x_0}\oplus \ad P_{x_1}\,\colon \ad_\psi(A_0)=A_1\}\subseteq (\bar\pi_0,\bar\pi_1)^*(\ad P\oplus\ad P)\subseteq (\bar\pi_0,\bar\pi_1)^*(\at P\oplus\at P)$

As a consequence we have identifications for the subspaces of $\bar\pi_0$ or $\bar\pi_1$-vertical tangent vectors, 
\begin{equation}\label{tartri}
\begin{aligned}
V^{\bar \pi_0}\End P&=(\bar\pi_0^*\ad P\oplus\bar\pi_1^*\at P)/\mathrm{Sym}\simeq \bar\pi_1^*\at P\\
V^{\bar \pi_1}\End P&=(\bar\pi_0^*\at P\oplus\bar\pi_1^*\ad P)/\mathrm{Sym}\simeq \bar\pi_0^*\at P
\end{aligned}
\end{equation}
(called target and source trivialisations), while for $V^{(\bar \pi_0,\bar\pi_1)}\End P$ both trivialisations exist, allowing two interpretations $V^{(\bar \pi_0,\bar\pi_1)}\End P\simeq \bar\pi_0^*\ad P$ or $V^{(\bar \pi_0,\bar\pi_1)}\End P\simeq \bar\pi_1^*\ad P$, related one to another by $(\psi,A_0)\in \bar\pi_0^*\ad P\simeq (\psi,-\ad_\psi(A_0))\in\bar\pi_1^*\ad P$, where $\ad_\psi\colon \ad P_{\bar\pi_0(\psi)}\rightarrow \ad P_{\bar\pi_1(\psi)}$ is the linear morphism induced by $\psi\colon P_{\bar\pi_0(\psi)}\rightarrow P_{\bar\pi_1(\psi)}$

\subsection{Reduced field theories on principal bundles}
The natural projections $\pi_J$ and $\pi^G$ determine an isomorphism $\pi_J\times\pi^G\colon JP\rightarrow P\times_X\con$, and the action $j\lambda$ on $JP$ is identified by this isomorphism with the action $\lambda\times \Id_{\con}$. For the particular case of a principal $G$-bundle, giving a lagrangian function $\La\colon JP\rightarrow \RR$ is equivalent to giving a function $\ell\colon (p,\chi)\in P\times_X\con\rightarrow \ell(p,\chi)\in\RR$, that we call the trivialized form of the Lagrangian. If the Lagrangian is invariant by the action of some closed sub-group $H\subseteq G$, this function can be expressed as $\bar\ell(Hp,\chi)$ for some function $\bar\ell\colon P/H\times_X\con$, a function that is called the $H$-reduced Lagrangian associated to $\La$.

An element in $P/H$ is an $H$-orbit. Sections of the bundle $P/H\rightarrow X$ can be identified with sub-bundles $Q\subset P$ where $H$ acts transitively, sub $H$-bundles of $P$ also called $H$-structures on the principal $G$-bundle $P$. The bundle $P/H\rightarrow X$ is therefore called the bundle of $H$-structures and denoted by $\pi_{\Str{H}}\colon \Str{H}\rightarrow X$. We also observe that $\pi^H\colon P\rightarrow P/H=\Str{H}$ itself has a natural structure of principal $H$-bundle (though on a base manifold that is not $X$). We shall denote this principal bundle as $P^{\Str{H}}$ (the same bundle space $P$, but a fibration on $\Str{H}$ instead of $X$, and with structure group $H\subseteq G$). When computing the partial derivatives (components in $\Str{H}$ and in $\con$) of the differential $\dd\bar\ell$ we must take into account that the natural projection $\pi^H\colon P\rightarrow \Str{H}$ determines a commutative diagram of exact sequences of vector bundles on $\Str{H}$:
\begin{equation}
\xymatrix{
	0\ar[r] & \ad P^{\Str{H}} \ar@{^{(}->}[r] \ar[d]^{0} &
	\pi_{\Str{H}}^*\at P \ar[d]^{\overline{\dd\pi}}\ar[r]^{\overline{\dd\pi^H}} & T\Str{H} \ar[d]^{\dd\pi_{\Str{H}}} \ar[r]& 0\\
	0\ar[r] & 0 \ar@{^{(}->}[r] &\pi_{\Str{H}}^*TX \ar[r]^{\Id} & \pi_{\Str{H}}^*TX\ar[r] & 0
}
\end{equation}
where $P^{\Str{H}}$ stands for the principal $H$-bundle $\pi^H\colon P\rightarrow \Str{H}$, whose $\pi^H$-vertical $H$-invariant vector fields on any orbit $Hp\in\Str{H}$ naturally extend to a $\pi$-vertical $G$-invariant vector field on the whole orbit $Gp\in P/G=X$, defining the immersion $\ad P^{\Str{H}}\hookrightarrow \pi_{\Str{H}}^*\ad P\subset \pi_{\Str{H}}^*\at P$. In particular the bundle of $\pi_{\Str{H}}$-vertical vectors $V^{\pi_{\Str{H}}}\Str{H}\rightarrow \Str{H}$ can be identified with $\pi_{\Str{H}}^*\ad P/\ad P^{\Str{H}}$.

Finally we also observe that the groupoid $(\bar\pi_0,\bar\pi_1)\colon \End P\rightarrow X\times X$ has a groupoid action on $\pi_{\Str{H}}\colon \Str{H}\rightarrow X$, defined by $ac_{\psi}\colon Hp\in\Str{H}_{x_0}\mapsto Hp\psi\in \Str{H}_{x_1}$ whenever $\pi(p)=\bar\pi_0(\psi)=x_0$, $\bar\pi_1(\psi)=x_1$.

Any $H$-invariant lagrangian $\La$ on $JP$ is determined by some $H$-reduced Lagrangian $\bar\ell$ on $\Str{H}\fp{X} \con$. Critical sections for $\La$ are characterized by the annihilation $\EL(p)=0\in\Gamma(p^*V^*P)$ of the Euler-Lagrange tensor (\ref{ELeq}). We may use the identification $VP\simeq P\times_X\ad P$ to identify $p^*VP$ with $\ad P$, for any field $p\in\Gamma(P)$. Thus the Euler-Lagrange tensor $\EL(p)\in\Gamma(p^*V^*P)$ of a smooth Lagrangian $\La$ can be expressed as $\EL^{\ad}(p)\in\Gamma(\ad^* P)$, the trivialized form of this tensor. In the particular case of a Lagrangian $\La$ obtained from an $H$-reduced Lagrangian $\ell$,  Euler-Lagrange equations for a section $p\in\Gamma(P)$ for the Lagrangian $\La$ become equivalent to a system of Euler-Poincaré equations for the reduced field $(q,\chi)\in\Gamma(\Str{H}\times \con)$ determined by $q=Hp\in P/H=\Str{H}$, $\chi=G(jp)\in JP/G=\con$:
\begin{theorem}[\cite{CasGarRod13}]\label{thmrelevante}
	Let $\pi\colon P\rightarrow X$ be a principal $G$-bundle on a manifold $X$ with volume element $\vol_X$. Let $H\subseteq G$ be a closed subgroup, and denote by $\pi^H\colon JP\rightarrow \Str{H}\fp{X}\con$ the natural quotient morphism. Let $\bar\ell\colon \Str{H}\fp{X}\con\rightarrow \RR$ be an $H$-reduced lagrangian function and $\La=\bar\ell\circ\pi^H\colon JP\rightarrow\RR$ the associated lagrangian function. \newline For any local field $p\in\Gamma(P)$, and for the induced local $H$-reduced field $\pi^H\circ jp=(q,\chi)\in \Gamma(\Str{H}\fp{X}\con)$ there holds
	\begin{equation}\label{admisibility}
	\Curv\chi=0,\qquad \dd^\chi q=0
	\end{equation}
	that is, the induced principal connection $\chi$ is flat (the curvature vanishes) and the $H$-structure is $\chi$-parallel.
	Moreover, the following are equivalent:
	\begin{enumerate}
		\item The local field $p$ is critical for the variational problem with fixed boundary variations and lagrangian function $\La$.
		\item The local field $p\in\Gamma(P)$ satisfies the system of Euler-Lagrange equations $0=\EL(p)$ for $\La\colon JP\rightarrow \RR$, where $\EL(p)$ is described by (\ref{ELeq}).
		\item The local $H$-reduced field is critical for a constrained variational problem described on $\Str{H}\fp{X}\con$ by the constraints (\ref{admisibility}) and 0-order lagrangian function $\bar\ell\colon \Str{H}\fp{X}\con\rightarrow \RR$.
		\item The local $H$-reduced field $(q,\chi)\in\Gamma(\Str{H}\fp{X}\con)$ satisfies the system of Euler-Poincaré equations $0=\EP(q,\chi)$ for $\bar\ell\colon \Str{H}\fp{X}\con \rightarrow \RR$, where $\EP(q,\chi)$ is described by:
		\begin{equation*}
		\EP(q,\chi)=\Div_{\chi}\left(\parcial{\bar\ell}{\chi}(q,\chi)\right) -\mathcal{P}_q^*\left(\parcial{\bar\ell}{q}(q,\chi)\right)\in\Gamma(\ad^* P)
		\end{equation*}
	\end{enumerate}
\end{theorem}
In this result $\partial\bar\ell/\partial\chi\in\Gamma(\chi^*V^*\con)\simeq \Gamma(\ad^* P\otimes TX)$ (recall that $\con$ is an affine bundle on $T^*X\otimes \ad P$), $\partial\bar\ell/\partial q\in\Gamma((\ad P/q^*\ad P^{\Str{H}})^*)$ (recall that $q^*V\Str{H}\simeq \ad P/q^*\ad P^{\Str{H}}$), $\mathcal{P}_q^*$ is the natural immersion $\Gamma((\ad P/q^*\ad P^{\Str{H}})^*)\subset \Gamma(\ad^* P)$ as the subspace of linear forms that vanish on $q^*\ad P^{\Str{H}}\subset \ad P$, and finally the divergence operator $\Div_{\chi}\colon \Gamma(\ad^*P\otimes TX)\rightarrow \Gamma(\ad^*P)$ associated to $\chi$  represents the differential operator adjoint to $\dd^\chi$, in the sense:
\begin{equation*}
\Div_{\chi}(\theta)\otimes \vol_X=\dd^\chi\left( i_\theta \vol_X\right)
\end{equation*}
here $i_\theta\vol_X$ stands for the $\ad P$-valued differential $(n-1)$-form obtained by contraction of $\theta$ with the volume element $X$ and $\dd^\chi$ stands for the covariant differential on alternating forms with values on $\ad P$ and on $\ad^*P$ obtained as extension of $\dd^\chi\colon \Gamma(\ad P)\rightarrow \Gamma(\ad P\otimes T^*X)$. We should observe that the flatness and parallelism conditions appearing in this theorem are necessary and sufficient condition for the section $(q,\chi)$ to be the $\pi^H$-projection of some local section $p\in\Gamma(P)$ (we also say that $p$ is a representation of the field $(q,\chi)$ in terms of potentials).

As a remark we should indicate that we use the natural identification $(JP)/H\simeq  \Str{H}\fp{X} \con$, as used in \cite{CasGarRod13}. An alternative method \cite{ElliGayHolmRati11} to describe the reduced space takes an specific principal connection on the bundle $P^H$ to establish the identification $(JP)/H=J(P/H)\fp{\Str{H}} (T^*X\otimes \ad P^H)$. Therefore, in a noncanonical way, depending on a connection choice, one has $ \Str{H}\fp{X} \con\simeq J\Str{H}\fp{\Str{H}} (T^*X\otimes \ad P^H)$ (fibered product of bundles over $\Str{H}$). With the situation described in \cite{CasGarRod13} one assumes that the configuration bundle is a principal $G$-bundle, and considers a subgroup $H\subset G$ of symmetries for the Lagrangian. The other choice \cite{ElliGayHolmRati11} needs not to assume that $P\rightarrow X$ is a principal bundle, only that there is a symmetry group $H$ with a proper, free action on the fibers and inducing the principal $H$-bundle $P^H\rightarrow \Str{H}=P/H$, using then some choice of principal connection on $P^H$ to perform the reduction.

Considering now the bundle $\pi\colon P\rightarrow X\subset \RR^2$ of admissible rod motions, in the hyperelastic case its dynamics is given by some trivialized Lagrangian with three components $\bar\ell=\Sigma-\Omega-\omega$. The kinetic-energy component can be written as $\Sigma(p,\chi)=\frac12 \|\tau-\hat\tau(x)\|_K^2$, where $\tau=\langle\chi,\partial/\partial t\rangle\in \at P$, it satisfies $\overline{\dd\pi}(\tau)=\partial/\partial t\in TX$, and determines a squared distance  $\|\tau-\hat\tau\|_K$ when  compared with a reference element $\hat\tau\in(\overline{\dd\pi})^{-1}(\partial/\partial t)\subset \at P$, using some positive-definite metric $K$ on the vector bundle $\ad P$. The strain-energy component can be written as $\Omega(p,\chi)=\frac12 \|\sigma-\hat\sigma(x)\|_W^2$ where $\sigma=\langle\chi,\partial/\partial s\rangle\in\at P$ also satisfies $\overline{\dd\pi}(\sigma)=\partial/\partial s\in TX$, and we have some positive-definite metric $W$ on $\ad P$ and a fixed  $\hat\sigma\in(\overline{\dd\pi})^{-1}(\partial/\partial s)\subset\at P$. Finally the potential-energy component can be written as $\omega(Hp)$ where $Hp=\pi_{Hp}(p,\chi)$ is determined using the natural projection $\pi_{Hp}\colon p\in P\times_X \con\mapsto Hp\in P/H$ and $\omega$ is some smooth function on $\Str{H}$.
\begin{equation}\label{disredlagrod}
\bar\ell(Hp,\chi)=\frac12 \|\tau-\hat\tau\|_K^2-\frac12 \|\sigma-\hat\sigma\|_W^2-\omega(Hp),\qquad \tau=\langle\chi,\partial/\partial t\rangle,\quad \sigma=\langle\chi,\partial/\partial s\rangle
\end{equation}
Let us observe again that the definition of the kinetic and strain energy components of the hyperelastic rod reduced lagrangian density rely on the choice of two sections $\hat\sigma,\hat\tau$ of Atiyah's bundle that project to $\partial/\partial s,\partial/\partial t$ respectively. Equivalently, this implies a choice $\hat\chi=\dd s\otimes \hat\sigma+\dd t\otimes \hat\tau$ of a principal connection. The hyperelastic rod Lagrangian is characterized by a connection $\hat\chi$ that describes the minimum energy configuration, a function $\omega$ that describes the potential of field forces acting on reduced configurations, and two positive-definite metrics $K,W$ on $\ad P$ that represent respectively the inertial and elastic properties (both linear and angular ones) of the rod.

\section{Discrete bundle theory on simplicial complexes}
A discrete bundle is a smooth bundle $\pi_d\colon Y_d\rightarrow V$ on a set $V$ (that is, $V$ is a $0$-dimensional manifold with discrete topology and all fibers $Y_{dv}=\pi_d^{-1}(v)$ are nonempty smooth manifolds). A section $(y_{dv})_{v\in V}$ shall be called a discrete field or field of vertex configurations. When we interpret $V$ as a $0$-dimensional, totally disconnected manifold, we directly obtain a notion of vertical bundle $\pi_{VY_d}\colon VY_d\rightarrow Y_d$ (whose fiber at $y_{dv}\in Y_{dv}$ is the tangent space $T_{y_{dv}}Y_{dv}$ of the fiber). Observe that the fiber associated to any element $y_{dv}$ is open and closed in $Y_d$, therefore taking the tangent space of the whole manifold $Y_d$ or of the fiber $Y_{dv}$ is the same $T_{y_{dv}}Y_d=V_{y_{dv}}Y_d=T_{y_{dv}}Y_{dv}$. In the case of discrete bundles we shall use the notion of vertical bundle and always avoid the notion of tangent bundle (because both of them are coincident).

Moreover the notion of tangent space of the base manifold becomes trivial (the tangent space of a 0-dimensional manifold $V$ is a bundle with trivial fiber $\{0\}$) and with the smooth notion of lagrangian density (an horizontal $n$-form), any  lagrangian density would be null. As a consequence smooth action functionals are identically zero and the whole smooth variational theory is trivial for discrete bundles. We may however define a discrete counterpart of variational principles for the case of discrete fields (sections of a bundle on a discrete space).

A common situation where discrete fields are considered is the discretization of a mechanical system described by trajectories $q(t)$ (hence sections of $Q\times T\rightarrow T$ for some time interval $T\subset\RR$ and configuration space $Q$) \cite{ChriMuntOwre11,CortMart01,FernGarcRodr04,HaiLubWan06,IseMunNorZan05,McLa,McLaQuis06,LeonDiegSant04,MarsPekaShko00,MarsWest01}. In this case dynamics is governed by some variational principle given through a lagrangian density $\La(t,q(t),\dot q(t))\dd t$. To discretise this theory we fix a grid of time points $t_0<t_1<\ldots<t_k$, an immersion $t\colon J\rightarrow T$ from a discrete interval $J\subset\ZZ$ into $T\subset\RR$, and we consider sequences $(q_j)_{j\in J}$ as approximate values for $q(t_j)$, where  $q(t)$ is a smooth trajectory. Hence discrete trajectories can be seen as sections of the discrete bundle $Q\times J\rightarrow J$. The action functional $\mathbb{L}\colon q(t)\mapsto \int_T \La(t,q(t),\dot q(t))\dd t$ on an interval $T\subset\RR$  is approximated through $\sum_{(j,j+1)\in J^1} L_j(q_j,q_{j+1})\cdot h_j$ where $J^1$ is the family of edges (pairs of consecutive elements $(j,j+1)$) in the discrete segment $J$ and $h_j$ is the length of the interval $[t_j,t_{j+1}]$. The discrete lagrangian function $L_j(q_j,q_{j+1})\cdot h_j$ is assumed to be a good approximation for $\int_{[t_j,t_{j+1}]}\La(t,q(t),\dot q(t))\dd t$. Taking for example $Q$ to be an affine space, a simple quadrature rule leads to the family of functions $L_j(q_j,q_{j+1})\cdot h_j=\La(t_j,q_j,\frac{q_{j+1}-q_j}{h_j})\cdot h_j$, determined from the immersion $t$, the smooth Lagrangian $\La$, and a linear structure on $Q$. A necessary condition for a discrete trajectory to be critical for this action functional is:
\begin{equation*}
D_2L_{j-1}(q_{j-1},q_{j})h_{j-1}+D_1L_{j}(q_j,q_{j+1})h_j=0
\end{equation*}
known as discrete Euler-Lagrange equations associated to the lagrangian density $(L_jh_j)_{(j,j+1)\in J^1}$. Here $D_1$, $D_2$ stand for the two partial derivatives, that is, the two components of $\dd L=(D_1L,D_2L)\in T^*Q\oplus T^*Q$.

These discretization techniques use two fundamental assumptions:  a topological notion of intervals $(j,j+1)\in J^1$ determined by neighboring discrete temporal events is needed to formulate a variational principle on discrete trajectories. On the other hand some immersion $t\colon J\hookrightarrow T$ together with a mechanism to compute differences $q_{j+1}-q_j$ of any pair of configurations are not essential to the formulation of a discrete variational principle but are needed if we want to formulate a discrete counterpart of a smooth lagrangian density. In the same manner also in the variational formulation of discrete field theories a certain topological structure will be needed for the formulation of a discrete variational principle. Additional structures can be used to obtain discrete principles using some smooth lagrangian density.

The formulation of an action functional and of a discrete variational principle, that in the smooth case is obtained by integration of a horizontal volume form (lagrangian density) along the jet extension of a section, should be introduced by other means in discrete field theories. On a discrete space $V$ a notion of integration can be achieved if we endow this space with a cellular complex structure \cite{CasiRodr12,CasiRodr12b}. With this structure, densities may be viewed as co-chains, integration domains as chains, and integration of a density on a domain as the natural pairing of a chain with a co-chain. We shall next introduce a particular model of discrete space, the notion of $n$-dimensional abstract simplicial complex.

\begin{define}
	For any set $V$ we denote by $V^{\times n}$ the product of $n+1$ copies of the set. Ordered sequences $(v_0,\ldots, v_n)\in V^{\times n}$ without repeated terms shall be called (ordered) abstract $n$-dimensional simplices on $V$.\newline We call abstract $n$-dimensional simplicial complex structure on a set $V$ any choice of a subset of abstract $n$-dimensional simplices $V^n\subset V^{\times n}$, that we call set of (ordered) facets of the simplicial complex. Any subsequence (preserving the order) $\alpha\subseteq \beta$ with $r+1$ elements of a facet $\beta\in V^n$ is called $r$-dimensional simplex of the simplicial complex. The set of $r$-dimensional simplices of the simplicial complex is denoted by $V^r\subset V^{\times r}$. The disjoint union $\mathcal{V}=V^0\sqcup V^1\sqcup\ldots\sqcup V^n$ is called the set of simplices of the simplicial complex. Elements in $V^1$ are called edges of the simplicial complex, and elements in $V^0$ are called vertices of the simplicial complex.
\end{define}
Observe that any of the terms $v\in V$ of a given simplex $\alpha\in V^r$ in the simplicial complex is necessarily a vertex $v\in V^0$ of the simplicial complex. There is no problem to assume that the simplicial complex structure is constructed on the set $V=V^0$.

\begin{define}
	On the space of simplices $\mathcal{V}$ there exists a natural topology: We say a simplex $\alpha\in\mathcal{V}$ is adherent to a simplex $\beta\in\mathcal{V}$ and write $\alpha\prec\beta$ if $\alpha$ is a subsequence of $\beta$. Whenever $\alpha$ is adherent to $\beta$, we shall say that $\beta$ contains $\alpha$.
	
	We call closure $\Cl K\subset V^0$ of a subset of cells $K\subset \mathcal{V}$ the set of all vertices adherent to some $\alpha\in K$.
	
	We call star $\Star^n_K\subset V^n$ of a subset of cells $K\subset \mathcal{V}$ the set of all facets containing some $\alpha\in K$.
	
	For any subset of facets $K\subset V^n$ we call interior vertex $v$ any vertex for which $\Star^n_v\subset K$ holds. We call exterior vertex any vertex for which $\Star^n_v\cap K=\emptyset$. We call boundary vertex any vertex that is not interior or exterior. Correspondingly, the set of interior, exterior and frontier vertices are denoted by $\Int K$, $\Ext K$, $\Fr K$.
	
	We say the simplicial complex is locally finite if the star of any of its vertices $v\in V^0$ is a finite set of facets. 
\end{define}
As the closure of a finite subset is a finite subset of vertices, if the simplicial complex is locally finite, also the star of a finite subset is a finite subset of facets.

For any integer $0\leq k\leq r$ there exists a projector $\pi_k\colon V^{\times r}\rightarrow V$, whose restriction to $V^r$ shall be denoted by $\pi_k\colon V^r\rightarrow V$. For any $r$ and any monotone integer sequence $k=(0\leq k_0<k_1<\ldots<k_s\leq r)$ there exists a corresponding projection $(\pi_{k_0},\ldots,\pi_{k_s})\colon V^r\rightarrow V^{\times s}$, whose image is in $V^s$, therefore defining projectors $\pi_k\colon V^r\rightarrow V^s$. For each $\alpha\prec\beta\in V^r$ we call $k(\alpha,\beta)$ the unique monotone sequence $k=(0\leq k_0<k_1<\ldots<k_s\leq r)$ such that $\pi_k(\beta)=\alpha$ (uniqueness of this sequence $k(\alpha,\beta)$ is ensured by the condition of non-repeated vertices that was assumed in the definition of abstract simplex). If $\beta=(v_0,v_1,\ldots, v_r)$ and $k(\alpha,\beta)=(0\leq k_0<k_1<\ldots<k_s\leq r)$ then $\alpha=(v_{k_0},v_{k_1},\ldots, v_{k_s})$.

Consider a simplicial complex $\mathcal{V}$ modeled on the set of vertices $V^0=V$. Consider any discrete bundle $\pi_d\colon Y_d\rightarrow V$ on these vertices.
We call space of $r$-dimensional simplex configurations the  bundle $\pi^r_d\colon Y^r_d\rightarrow V^r$ obtained as the restriction of $\pi_d^{\times r}\colon Y^{\times r}_d\rightarrow V^{\times r}$ to the set of $r$-dimensional simplices $V^r\subset V^{\times r}$. There exist obvious projectors $\pi_k\colon Y^r_d\rightarrow Y^s_d$ for any monotone interger sequence $k=(0\leq k_0<k_1<\ldots<k_s\leq r)$, with the property $\pi^s_d\circ \pi_k=\pi_k\circ \pi^r_d$. For any $\alpha\prec\beta$ we denote $\pi_\alpha^\beta\colon Y^r_{d\beta}\rightarrow Y^s_{d\alpha}$ the restriction to $Y^r_{d\beta}$ of the projector $\pi_k$ with $k=k(\alpha,\beta)$.

\begin{define}
	For any discrete bundle $\pi_d\colon Y_d\rightarrow V$ on the space of vertices of an $n$-dimensional simplicial complex $\mathcal{V}$, we call discrete jet bundle the bundle $j\pi_d=\pi_d^n\colon JY_d=Y^n_d\rightarrow V^n$. Any of its elements is called a facet configuration or a discrete jet.
\end{define}
A discrete jet of the bundle $\pi_d\colon Y_d\rightarrow V$ is then a specific choice of facet $\beta=(v_0,v_1,\ldots, v_n)\in V^n$, together with a sequence $(y_0,y_1,\ldots y_n)$ of configurations $y_k\in Y_{dv_k}$ on each of its adherent vertices $v_k=\pi_k(\beta)$. Using the discrete bundles $\pi_k^*Y_d\rightarrow V^n$ determined by the projector $\pi_k\colon V^n\rightarrow V$, the discrete jet bundle $JY_d$ can be seen as the fibered product (over $V^n$) of the bundles $\pi_0^*Y_d,\pi_1^*Y_d,\ldots, \pi_n^*Y_d$. Using the projections to each component $\pi_{k}\colon JY_d\rightarrow Y_d$ we may then state $VJY_d=\bigoplus_{k=0}^n\pi_k^*VY_d$ obtaining immersions $i_k\colon \pi_k^*VY_d\hookrightarrow VJY_d$.

For any section $y_d\in\Gamma(Y_d)$, we call jet extension $jy_d=y^n_d\in\Gamma(JY_d)$ the section whose components are $(y_d\circ\pi_k)\in\Gamma(\pi_k^*Y_d)$. In a similar way, for any vertical field $\delta y_d\in\Gamma(y_d^*VY_d)$ defined along a section $y_d$, we call jet extension $j\delta y_d=\delta y^n_d\in\Gamma((jy_d)^*VJY_d)$ the section whose components are $(\delta y_d\circ\pi_k)$.

\begin{define}
	We call discrete lagrangian density any smooth function $L_d\colon JY_d\rightarrow \RR$ on the discrete jet bundle. A discrete lagrangian density determines then a family of smooth functions $L_{d\beta}\colon JY_{d\beta}=Y_{dv_0}\times Y_{dv_1}\times \ldots\times Y_{dv_n}\rightarrow \RR$, one for each facet $\beta=(v_0,v_1,\ldots, v_n)\in V^n$.
\end{define}
We denote by $\dd^kL_{d}\in \Gamma(\pi_k^*V^*Y_d)$ the corresponding components of its differential $\dd L_{d}$, using the inclusion $i_k\colon \pi_k^*VY_d\hookrightarrow VJY_d$.

\begin{define}
	Consider a discrete lagrangian density $L_d\colon JY_d\rightarrow \RR$ defined for a discrete bundle $\pi_d\colon Y_d\rightarrow V$ on the space of vertices $V$ of an $n$-dimensional locally finite simplicial complex $\mathcal{V}$. We call action functional associated to $L_d$ and to any domain (finite family of facets $K\subset V^n$) the following:
	\begin{equation*}
	\LL_K(y_d)=\sum_{\beta\in K} L_{d\beta}(jy_{d\beta}) 
	\end{equation*}
	defined for sections $y_d\in\Gamma(\Cl K,Y_d)$.
\end{define}
As $\Cl K$ contains a finite number of vertices, $\Gamma(\Cl K,Y_d)=\prod_{v\in \Cl K} Y_{dv}$ is a finite-dimensional manifold.  The differential of $\LL_K$ can be decomposed as a direct sum of components, each associated to a vertex $v\in\Cl K$. A specific element of this manifold is a critical point for the smooth function $\LL_K$ precisely when:
\begin{equation*}
0=\sum_{\beta\in K\cap \Star^n_v} \dd^{k(v,\beta)}_{jy_{d\beta}}L_{d\beta}\in V^*_{y_d(v)}Y_{d},\qquad \forall v\in \Cl K
\end{equation*}
however the functional $\LL_K$ may not be bounded, this condition is usually too strong and critical points in this sense may not exist. In the case of vertices interior to $K$, we know that $\Star^n_v\subset K$ therefore a weaker necessary condition is:
\begin{equation}\label{discreteEL}
0=\EL_{dv}(y_d)=\sum_{\beta\in \Star^n_v} \dd^{k(v,\beta)}_{jy_{d\beta}}L_{d\beta}\in V^*_{y_d(v)}Y_d,\qquad \forall v\in \Int K
\end{equation}
In this case we say the point is critical for $\LL_K$ for fixed boundary variations (variations vanishing at the boundary $\Fr K$). The section $\EL_d(y_d)\in\Gamma(y_d^*V^*Y_d)$ is a discrete analogue to (\ref{ELeq}) and is called the Euler-Lagrange tensor associated to the discrete lagrangian density $L_d$ and discrete field $y_d\in\Gamma(Y_d)$.

Any vertical vector field $D\in\Gamma(Y_d,VY_d)$ on $Y_d$ determines a corresponding vector field $D^{\times n}=(D,\ldots, D)$ on $Y_d^{\times n}$. Its restriction to $JY_d\subset Y^{\times n}_d$ shall be denoted by $jD$, and called extension of $D$ to the discrete jet bundle $JY_d$.
\begin{define}
	We say $D\in\Gamma(Y_d,VY_d)$ is an infinitesimal symmetry for a discrete lagrangian density $L_d\colon JY_d\rightarrow \RR$ if $jD(L_d)=0$.
\end{define}

Considering the value of $jD(L_d)$ at $jy_{d\beta}$ for any discrete field and any facet, we have
\begin{equation*}
\left(jD(L_d)\right)(jy_{d\beta})=\sum_{v\prec\beta}\langle \dd^{k(v,\beta)}_{jy_{d\beta}}L_{d\beta},D(y_{dv})\rangle
\end{equation*}
which vanishes if $D$ is an infinitesimal symmetry.
\begin{define}
	Consider a discrete bundle $Y_d\rightarrow V$ on the set of vertices of an $n$-dimensional simplicial complex, a discrete Lagrangian $L_d\colon JY_d\rightarrow V^n$, and any finite domain $K\subset V^n$. We call discrete Noether current $\mu_{dK}(y_d)\in\Gamma(y_d^*V^*Y_d)$ associated to $K$ and to some discrete field $y_d\in\Gamma(Y_d)$ the section given,  at any vertex $v$, as:
	\begin{equation*}
	\left(\mu_{dK}(y_d)\right)(v)=\sum_{\beta\in K\cap \Star^n_v} \dd^{k(v,\beta)}_{jy_{d\beta}}L_{d\beta}
	\end{equation*}
\end{define}
We observe that $\mu_{dK}(y_d)$ vanishes at $v\in\Ext K$ (because $K\cap\Star^n_v=\emptyset$). Moreover if the discrete field $y_d$ is critical it also vanishes at $v\in \Int K$ (because $K\cap \Star^n_v=\Star^n_v$ and the definition of $\mu_{dK}(y_d)$ at $v$ in this case is coincident with $\EL_{dv}(y_d)$). The discrete Noether current associated to a finite domain $K$ and a critical discrete field $y_d$ has support on $\Fr K$. There is a natural bilinear product (integration on the boundary) $\langle \mu,\delta y_d\rangle=\sum_{v\in\Fr K} \langle \mu(v),\delta y_d(v)\rangle\in\RR$ for sections $\mu\in\Gamma(\Fr K,y_d^*V^*Y_d)$ and  $\delta y_d\in\Gamma(y_d^*VY_d)$ that determines a real value, called total Noether current associated to $\delta y_d\in\Gamma(y_d^*VY_d)$.
\begin{theorem}[Discrete Noether conservation law]\label{Noether}
	If $\mu_{dK}(y_d)\in\Gamma(y_d^*V^*Y_d)$ is the discrete Noether current associated to $K$ and to some critical discrete field $y_d\in\Gamma(Y_d)$ and $\delta y_d=D\circ y_d\in\Gamma(y_d^*VY_d)$ is the restriction to $y_d$ of an infinitesimal symmetry $D\in\Gamma(Y_d,VY_d)$ of the discrete Lagrangian $L_d$, then the total Noether current associated to $\delta y_d$ vanishes.
\end{theorem}
PROOF:\newline	As $D$ is an infinitesimal symmetry, for each facet $\beta\in V^n$, we know that $jD_\beta$ is incident with  $\dd L_{d\beta}$.  We get $0=\sum_{v\prec\beta} \langle\dd^{k(v,\beta)}_{jy_{d\beta}}L_{d\beta},D(y_d(v))\rangle$. Addition of this expression on all facets $\beta\in K$ leads to:
\begin{equation*}
\begin{aligned}
0&=\sum_{\beta\in K}\sum_{v\prec\beta} \langle\dd^{k(v,\beta)}_{jy_{d\beta}}L_{d\beta},D(y_d(v))\rangle
=\sum_{v\in\Cl K}\sum_{\beta\in K\cap\Star^n_v} \langle\dd^{k(v,\beta)}_{jy_{d\beta}}L_{d\beta},D(y_d(v))\rangle=\\
&=\sum_{v\in\Cl K}\langle \mu_{dK}(y_d)(v),D\circ y_d(v)\rangle
=\sum_{v\in\Fr K}\langle \mu_{dK}(y_d)(v),D\circ y_d(v)\rangle
\end{aligned}
\end{equation*}
the last equality because the Noether current has support $\Fr K$ if we know that $y_d$ is critical.$\square$

The bilinear product that computes the total current is the addition at all vertices $v\in\Fr K$ of the real values $\langle \mu_{dK}(y_{d})(v),\delta y_{d}(v)\rangle$, and may be seen as the integration of a density on the boundary of the domain $K$. This represents a discrete analogue of Noether's conservation principle for currents determined from a vertical infinitesimal symmetry of a smooth lagrangian density.

\begin{remark}\label{remark01}
	The Euler-Lagrange tensor or Noether current associated to some discrete Lagrangian are created here in an independent way from what is done for a smooth Lagrangian. In the case that we generate discrete Lagrangians from a smooth one, certain questions arise, namely, assuming that there is a measure of error for a discrete field when it is considered as approximation of a smooth field (for fields on $Y_d$ compared to fields on $Y$, fields on $VY_d$ compared to fields on $VY$ and so on), one may ask if there exists any bound for the corresponding error of the value of the discrete action functional on the discrete field, with respect to the value of the smooth action functional on the smooth field, and similar questions arise for the corresponding smooth/discrete Euler-Lagrange tensors or conserved Noether currents associated to symmetries. In particular, formulating this last question one assumes that the discretization procedure transforms smooth Lagrangians with known infinitesimal symmetry $D$ into discrete Lagrangians that have the same infinitesimal symmetry. 
	
	Dealing with these issues, in their presentation of discrete mechanics, Marsden and West \cite{MarsWest01} defined an exact discrete Lagrangian $L^E_d$ associated to any smooth Lagrangian $L$. They observed that if another discrete Lagrangian $L_d$ approximates the exact discrete Lagrangian to order p, then the corresponding discrete Hamiltonian flow that solves the discrete Euler-Lagrange equations, is order p accurate when used as a one-step numerical scheme to approximate smooth solutions of Euler-Lagrange equations. Leok and Shingel \cite{LeoShi12}  develop a systematic method to construct discrete lagrangians (and variational principles) from smooth Lagrangians in mechanics, using a choice of a numerical quadrature formula, together with a choice of a finite-dimensional function space or a one-step numerical scheme. The order of the quadrature formula and numerical scheme determines the order of accuracy and momentum-conservation properties of the associated variational integrators. A systematic study is done for the case of covariant Lagrangian densities on a Lie group, using the exponential mapping, or the Cayley transform on $\mathrm{SO}(3)$ as retraction mapping (the latter improves the computational cost, keeping the order of accuracy), and fixing an arbitrary quadrature rule with several nodes. In \cite{VanLiaLeo12} it is suggested that this variational error analysis may be extended to discrete field theories, and that one may develop general techniques for constructing variational integrators for field theories, extending in this way the theory of variational error analysis introduced by \cite{MarsWest01} and explored in \cite{LeoShi12}, among others. Moreover \cite{VanLiaLeo12} also remarks that a corresponding theory of variational error analysis for the discretization of field theories may also rely on a deeper understanding of the associated boundary Lagrangians and how they serve as generating functionals for multisymplectic relations.
	
	Following this order of ideas, it makes sense the to explore the possibility to discretise a smooth lagrangian density when we fix a numerical cubature rule to approximate integrals in n-dimensional domains and fix a one-step numerical scheme to approximate solutions of certain partial differential equations. The order of accuracy of both tools should be reflected into the order of precision of discrete critical fields of the discretised variational principle when used as approximation of the smooth critical fields of the smooth variational principle. Such a comparison should be performed using the exact discrete Lagrangian as reference. That is, discrete fields that are critical for any discretised Lagrangian should be compared to discrete fields that are critical for the exact discrete Lagrangian, and these should be compared to smooth fields that are critical for the original smooth variational principle.
	
	In the remaining we explore the mentioned possibility, in a framework that respects symmetries and reduction processes, but don't elaborate the corresponding error analysis, which demands tools on order of precision for cubature rules and numerical schemes for PDEs that are off-topic for our present objectives. 
\end{remark}
We enter now the particular case of a discrete principal $G$-bundle and determine a trivialization of this bundle in terms of discrete connections, in a similar way as was discussed for the smooth principal bundles.

Observe that any discrete bundle morphism $f\colon Y_d\rightarrow Z_d$ from the bundle $Y_d\rightarrow V$ to the bundle $Z_d\rightarrow V$ (projecting as $\Id_V$ on $V$) determines $f^{\times r}\colon Y^{\times r}_d\rightarrow Z^{\times r}_d$, whose restriction to $Y^r_d$ is a bundle morphism $f^r\colon Y^r_d\rightarrow Z^r_d$ (projecting as $\Id$ on $V^r$). In particular, when the group $G$ acts on $Y_d$ by $\lambda_g\colon Y_d\rightarrow Y_d$, it also acts on each bundle $Y^r_d$ using $\lambda^r_g$.

Consider a discrete principal $G$-bundle $\pi_d\colon P_d\rightarrow V$. From this discrete bundle, using $V$ as a $0$-dimensional manifold, there exists a corresponding notion of adjoint bundle $\pi_{d\ad}\colon \ad P_d=VP_d/G\rightarrow V$, and Gauge bundle $\pi_{d\Gau}\colon \Gau P_d\rightarrow V$. In particular, there still holds $VP_d\simeq P_d\times_V \ad P_d$ and hence the discrete Euler-Lagrange tensor $\EL_d$ or discrete Noether currents $\mu_{dK}$ can be expressed as sections $\EL^{\ad}_d,\mu_{dK}^{\ad}$ of $\ad^*P_d$ using the identification $p_d^*VP_d\simeq \ad P_d$.

However, as the base manifold is a totally disconnected 0-dimensional manifold, the notions of tangent bundle and Atiyah bundle coincide with the vertical and adjoint bundles. In a similar way the notions of jet bundle and connection bundle, in the sense of the smooth category, would be trivial bundles.

Moreover, the classical notion of Ehresmann bundle as $\End P_d=(P_d\times P_d)/G$ includes too many fiber-to-fiber endomorphisms that may not be compatible with the set $V^1$ of edges present in our simplicial complex structure.

\begin{define}
	For a discrete principal $G$-bundle $\pi_d\colon P_d\rightarrow V$ on the space of vertices of an $n$-dimensional simplicial complex, consider the extended bundle on edges $\pi^1_d\colon P_d^1\rightarrow V^1$ and the extended bundle on facets $\pi^n_d\colon P^n_d\rightarrow V^n$.
	
	We call discrete bundle of connections the quotient bundle $\pi_{d\con}=\bar\pi_d^n\colon \con_d=(JP_d)/G\rightarrow V^n$ of the bundle of facet configurations (discrete jet bundle), by the action group $G$. Any of its elements is called a discrete connection element.\newline   We call Ehresmann discrete bundle the quotient bundle $\pi_{d\End}=\bar \pi_d^1\colon\End^1P_d=P^1_d/G\rightarrow V^1$ of the bundle of edge configurations, by the action group $G$. Any of its elements is called a discrete parallel transport along a given edge.
\end{define}
Ehresmann's discrete bundle is a groupoid bundle with projectors $\bar\pi_0=\pi_0\circ\bar\pi^1_d$, $\bar\pi_1=\pi_1\circ\bar\pi^1_d$. There exists a natural groupoid action on $P_d$, on $VP_d$, on $\ad P_d$ and on $\Gau P_d$.

Projectors $\pi_{0k}\colon JP_d\rightarrow P^1_d$ factor as $\bar\pi_{0k}\colon \con_d\rightarrow \End^1 P_d$. For any discrete connection $G(jp_\beta)\in JP_{d\beta}/G$ on a facet $\beta\in V^n$ the images by $\bar\pi_{0k}$ $(k=1\ldots n)$ are fiber-to-fiber endomorphisms, all with the common source $\pi_0(\beta)$ and different targets $\pi_k(\beta)$. That is, the image by $(\bar\pi_{0k})_{k=1,\ldots,n}$ of any element in $\con_{d\beta}$ with $\beta=(v_0,v_1,\ldots,v_n)\in V^n$ lies in:
\begin{equation*}
\End^{n}P_{d\beta}=\left\{ (\psi_1,\ldots,\psi_n)\,\colon\bar\pi_0(\psi_k)=v_0,\bar\pi_1(\psi_k)=v_k,\,\forall k\right\}\subset (\End^1 P_d)^{\times n-1}
\end{equation*}
This space determines a bundle $\pi_{d\con}\colon \End^{n}P_d\rightarrow V^n$, that can be identified with $\con_d=JP_d/G$ using $(\bar\pi_{01},\ldots,\bar\pi_{0n})$. The projector $\bar\pi_{d0}=\pi_0\circ \pi_{\con_d}\colon \End^{n}P_d\rightarrow V$ shall be called the source mapping on $\con_d$ or on $\End^{n}P_d$.

A discrete connection can be seen both as a $G$-orbit of a sequence $(p_0,p_1,\ldots,p_n)$ of configurations for all vertices in a facet $(v_0,v_1,\ldots,v_n)$ or as a sequence $(\psi_1,\psi_2,\ldots,\psi_n)$ of fiber-to-fiber endomorphisms sharing a common source vertex $v_0$ that, together with all target vertices $(v_1,\ldots, v_n)$ determine a facet $(v_0,v_1,\ldots,v_n)$ of the simplicial complex.

For any ordered pair $k=(0\leq k_0<k_1\leq n)$ the projector $\pi_k\colon JP_d\rightarrow P^1_d$ factors as $\bar\pi_k\colon \con_{d}\rightarrow \End^1 P_{d}$. Moreover, for $jp_\beta=(p_0,p_1,\ldots,p_n)$ we observe that $\pi_{k_1k_2}(jp_\beta)=p_{k_1}^{-1}p_{k_2}=\left(p_0^{-1}p_{k_1}\right)^{-1} \left(p_0^{-1}p_{k_2}\right)=\pi_{0k_2}(jp_\beta)\circ (\pi_{0k_1}(jp_\beta))^{-1}$. Knowledge of the components $\pi_{0k}(jp_\beta)$ leads to the remaining components $\pi_{k_1k_2}(jp_\beta)$. Hence for the identification $\con_d\simeq \End^{n}P_d$, the projectors $\bar\pi_{k_1k_2}$ are simply $\bar\pi_{k_1k_2}(\psi_1,\ldots,\psi_n)=\psi_{k_1}^{-1}\psi_{k_2}=\psi_{k_2}\circ\psi_{k_1}^{-1}$, which is valid even for the case $k_1=0$, taking the convention $\psi_0=\Id\in\Gau P_{dv_0}$.
\begin{define}
	Consider $\pi_d\colon P_d\rightarrow V$ and the projector $\pi_0\colon V^n\rightarrow V$. We call trivialization of the discrete jet bundle $JP_d\rightarrow V^n$ the morphism:
	\begin{equation*}
	(\pi_0, \bar\pi_{01}\circ \pi^G,\ldots,\bar\pi_{0n}\circ \pi^G)\colon JP_{d}\rightarrow \pi_0^*P_{d}\times_{V^n} \End^{n}P_d
	\end{equation*}
\end{define}
This represents an isomorphism of discrete bundles on $V^n$. The action $\lambda^n_g$ with this discretization behaves as the identity on all Ehresmann's components and as $\lambda_g$ on the first component $P_d$. As a consequence, $JP_d/H\simeq \pi_0^*\Str{H}\times_{V^n}\End^nP_d$.

The trivialization of the discrete jet bundle allows, for a discrete principal bundle, to represent any discrete Lagrangian $L_d\colon JP_d\rightarrow \RR$ as a function $\ell_d\colon JP_{d}\rightarrow \pi_0^*P_{d}\times_{V^n} \End^{n}P_d\rightarrow \RR$, a function $\ell_d(p_0,\psi_1,\ldots,\psi_n)$ that we call the trivialized discrete Lagrangian. If the discrete Lagrangian is invariant by the action of the closed subgroup $H\subseteq G$ then it determines a reduced discrete Lagrangian $\bar\ell_d(Hp_0,\psi_1,\ldots,\psi_n)$ on the reduced jet space $JP_d/H\simeq \pi_0^*\Str{H}_d\times_{V^n}\End^nP_d$, where $\Str{H}_d=P_d/H$.

\begin{define}
	Consider a discrete principal $G$-bundle $\pi_d\colon P_d\rightarrow V$ and a closed subgroup $H\subseteq G$. We call bundle of $H$-reduced discrete jet configurations $\pi_{dRJP}\colon RJP_d\rightarrow V^n$ the fibered product $\pi_0^*\Str{H}\times_{V^n}\End^nP_d$. There exists a natural identification $JP_d/H\simeq RJP_d$. There exist projectors $\bar\pi_0\colon RJP_d\rightarrow \Str{H}_d$, $\bar\pi_{0k}\colon RJP_d\rightarrow \pi_0^*\End^1 P_d$, determining components $rjp_\beta^0\in\Str{H}_d$ and $rjp_\beta^{0k}\in\End^1P_d$ for any reduced configuration $rjp_d\in RJP_d$.
\end{define}

Using the source trivialisation to identify $V\End P_d\simeq \bar\pi_{d0}^*\ad P_d$ (in this case $\bar\pi_{d0}\colon \End P_d\rightarrow V$ the source mapping) and $V\Str{H}_d\simeq \pi_{d\Str{H}}^*\ad P_d/\ad P_d^{\Str{H}}$ we get an identification $$V(RJP_d)\simeq \bar\pi_0^*\left(\pi_{d\Str{H}}^*\ad P_d/\ad P_d^{\Str{H}}\right)\oplus \bigoplus_{k=1}^n \bar\pi_{d0}^*\ad P_d$$
where $\bar\pi_{d0}\colon RJP_d\rightarrow V$ represents the source mapping and $\bar\pi_0\colon RJP_d\rightarrow \Str{H}_d$ the projection to the first component.

Using the identification of the dual of a direct sum as the direct sum of duals, and the dual of a quotient space $E/F$ as a subspace of linear functions $\omega\in E^*$ that are incident with $F$, we get, for any reduced discrete jet $rjp_\beta=(q_0,\psi_1,\ldots,\psi_n)$ the immersion:
\begin{equation*}
V^*_{rjp_\beta}(RJP_d)\subseteq \bigoplus_{i=0,\ldots, n}\ad^* P_{dv_0}
\end{equation*}
as the subspace of elements $(\omega_0,\omega_1,\ldots, \omega_n)$ with $\omega_0\in \ad^* P_{dv_0}$ incident to $\ad P^{\Str{H}}_{dq_0}\subset \ad P_{dv_0}$ 

As a consequence, and considering the source mapping $\bar\pi_{d0}\colon RJP_d\rightarrow V$, there exists a natural immersion:
\begin{equation*}
V^*(RJP_d)\subset \bigoplus_{i=0}^n\bar \pi_{d0}^*\ad P_d
\end{equation*}
where the fiber at $(q_0,\psi_1,\ldots,\psi_n)\in RJP_d$ is characterized by sequences $(\omega_0,\omega_1,\ldots,\omega_n)$ in $\ad^*P_{dv_0}$  whose first term $\omega_0$ vanishes on $\ad P^{\Str{H}}_{dq_0}\subset \ad P_{dv_0}$.
\begin{define}
	Consider a (discrete Lagrangian) function $\bar\ell_d$ on the bundle of $H$-reduced discrete jets $\bar\pi_0\colon RJP_d\rightarrow V$. Consider the immersion 	$V^*(RJP_d)\subset \bigoplus_{i=0}^n\bar \pi_{d0}^*\ad P_d$ determined by the source trivialisation of the vertical bundle $V\End P_d$. We denote by $(\partial^0\bar\ell_d,\partial^{01}\bar\ell_d,\ldots, \partial^{0n}\bar\ell_d)$ the corresponding $n+1$ components of the differential $\dd\bar\ell_d$. All these components are sections of $\bar\pi_{d0}^*\ad^*P_d$
\end{define}
\begin{theorem}
	If $L_d\colon JP_d\rightarrow \RR$ is a discrete Lagrangian $L_d=\bar\ell_d\circ\pi^H$ determined by some $H$-reduced discrete Lagrangian $\bar\ell(q_0,\psi_1,\ldots,\psi_n)$ on $RJP_d$, 	
	then the associated discrete Euler-Lagrange tensor $\EL_d(p_d)\in\Gamma(p_d^*V^*P_d)$ associated to $L_d$ and any section $p_d\in\Gamma(P_d)$ is identified by $p_d^*V^*P_d\simeq \pi_d^*\ad P_d$ with its trivialized version:
	\begin{equation}\label{dEP}
	\EL^{\ad}_{dv}(p_d)=\sum_{\pi_0(\beta)=v} \partial^0_{rj_{d\beta}}\bar\ell_\beta+
	\sum_{i=1}^n\sum_{\pi_0(\beta)=v}\partial^{0i}_{rj_{d\beta}}\bar\ell_\beta-
	\sum_{i=1}^n\sum_{\pi_i(\beta)=v}\ad^*_{rj_{d\beta}^{0i}}\partial^{0i}_{rj_{d\beta}}\bar\ell_\beta\in \ad^*P_{dv}
	\end{equation}
	where $rj_d\in \Gamma(RJP_d)$ is the natural extension of $p_{d}\in\Gamma(P_d)$, with components $rj_{d\beta}^0\in \Str{H}_{dv_0}$, $rj_{d\beta}^{0k}\in\End^1 P_{dv_0v_k}$ and $\ad_\psi^*\colon \ad^* P_{d\bar\pi_0\psi}\rightarrow \ad^* P_{d\bar\pi_1\psi}$ is the transpose linear morphism corresponding to $\ad_{\psi^{-1}}\colon \ad P_{d\bar\pi_1\psi}\rightarrow \ad P_{d\bar\pi_0\psi}$, induced on vertical vector fields by $\psi^{-1}\colon P_{d\bar\pi_1\psi}\rightarrow P_{d\bar\pi_0\psi}$. 
\end{theorem}
This result shows that in order to solve the system of discrete Euler-Poincaré equations we may choose to work with reduced fields instead of working with ``potentials'' (fields $p_d\in\Gamma(P_d)$)
\begin{define}
	For a discrete principal $G$-bundle $\pi_d\colon P_d\rightarrow V$ and a closed subgroup $H\subseteq G$ we call $H$-reduced discrete field any pair $(q_d,\psi_d)$ formed by a section $q_d\in\Gamma(\Str{H}_d)$ of the discrete bundle of $H$-structures and a section $\psi_d\in\Gamma(\End^1 P_d)$ of the discrete Ehresmann bundle.
\end{define}
Any section $\psi_d\in\Gamma(\End^1 P_d)$ is determined if we give discrete parallel transports along each edge $\alpha\in V^1$, and shall be called a discrete connection. 
We may observe that any discrete field $p_d\in\Gamma(P_d)$ determines an $H$-reduced discrete field, given as $q_{dv}=Hp_{dv}$ for each $v\in V$, and a discrete connection, given as $\psi_{dv_0v_1}=p_{dv_0}^{-1}p_{dv_1}\in \End^1 P_{dv_0v_1}$ for each $(v_0,v_1)\in V^1$. If $(q_d,\psi_d)$ is an $H$-reduced field obtained from some globally defined field $p_d\in\Gamma(P_d)$, then its components always satisfy $\psi_{dv_0v_1}\psi_{dv_1v_2}=\psi_{dv_0v_2}$ and $q_{dv_0}\psi_{dv_0v_1}=q_{dv_1}$, when $(v_0,v_1),(v_1,v_2),(v_0,v_2)\in V^1$. We say then that the discrete connection $\psi_d\in\Gamma(\End^1 P_d)$ is flat and that the discrete $H$-structure $q_d$ is $\psi_d$-parallel.

In a similar way as in the condition of admissibility (\ref{admisibility}) appearing in the variational principle for reduced variables in theorem \ref{thmrelevante}, we may define for the discrete case:
\begin{define}
	An $H$-reduced field $(q_d,\psi_d)\in\Gamma(\Str{H}_d)\times\Gamma(\End^1 P_d)$ is called admissible if the discrete connection component $\psi_d$ is flat:
	\begin{equation}\label{dflat}
	\psi_{dv_0v_1}\psi_{dv_1v_2}=\psi_{dv_0v_2} \qquad \forall (v_0,v_1),(v_1,v_2),(v_0,v_2)\in V^1
	\end{equation}
	and the discrete $H$-structure component $q_d$ is $\psi_d$-parallel:
	\begin{equation}\label{dpara}
	q_{dv_0}\psi_{dv_0v_1}=q_{dv_1} \qquad \forall (v_0,v_1)\in V^1
	\end{equation}
\end{define}
We already know that discrete fields $p_d$ always lead to $H$-reduced fields $(q_d,\psi_d)$ that are admissible. A local converse is also true: an admissible $H$-reduced field $(q_d,\psi_d)$ is locally determined by some discrete field $p_d$. More precisely, for any facet $\beta$, it is possible to give a section $p_d\in\Gamma(\Cl \beta,P_d)$ that projects to $q_d\in \Gamma(\Cl\beta,\Str{H}_d)$ and $\psi_d\in\Gamma((\Cl\beta)^1,\End^1 P_d)$. For this it suffices to choose the source vertex $\pi_0(\beta)=v$, choose an arbitrary element $p_{dv_0}\in q_{dv_0}\subset P_{dv_0}$ (recall that elements $q_{dv}\in\Str{H}_d$ are $H$-orbits contained in $P_d$), and define $p_{dv_k}=p_{dv_0}\psi_{dv_0v_k}$ for the remaining vertices $v_k=\pi_k(\beta)$. By definition this $p_d$ projects to $\Str{H}_{dv_0}$ as $q_{dv_0}$, to $\End^1P_{dv_0v_k}$ as $\psi_{dv_0v_k}$. The condition of flatness ensures that $p_d$ also projects to $\End^1P_{dv_jv_k}$ as $\psi_{dv_jv_k}$, for each of the remaining edges $\alpha\prec\beta$, and the condition of parallelism ensures that $p_d$ also projects to $\Str{H}_{dv_k}$ as 
$q_{dv_k}$. The global equivalence of admissible $H$-reduced fields and fields obtained by reduction of a potential field $p_d$ is a question whose answer depends on the topological structure of the principal bundle (this remark is valid both in the smooth and in the discrete formulations).

The fibered product of $n$ copies of $\pi_0\colon V^1\rightarrow V$ can be seen as a subset in $V^{\times n}$, given by sequences $(v_0,v_1,\ldots,v_n)$ such that each $(v_0,v_k)\in V^1$. The bundle $\End^nP_d\rightarrow V^{\times n}$ can be seen as the fibered product $(\End^1 P_d)^{\times_V n-1}$ of $n$ copies of $\bar\pi_{d0}\colon \End^1P_d\rightarrow V$, restricted to $V^n\subset V^{\times n}$. Therefore any section $\psi_d\in\Gamma(\End^1 P_d)$ determines a section $\psi^{\times n-1}_d$ of $(\End^1 P_d)^{\times_V n-1}$, and its restriction to $V^n$, that we denote by $\psi^n_d\in\Gamma(\End^nP_d)$. 

As a consequence any $H$-reduced discrete field $(q_d,\psi_d)$ determines its reduced jet extension $rj_d=(q_d,\psi^n_d)\in\Gamma(RJP_d)$. We say that some admissible  $H$-reduced discrete field $(q_d,\psi_d)$ is critical for the Lagrangian $\bar\ell_d$ if the discrete Euler-Poincaré tensor (\ref{dEP}) vanishes.

\section{Discretization and integration for hyperelastic rods}

As a model for 2-dimensional discrete space $\mathcal{V}$ we shall use the Coxeter-Freudenthal-Kuhn (CFK) partition of the plane \cite{CasiRodr17b}, that has a natural generalization to higher dimensions. In the specific two-dimensional case, the space $V^2$ of facets is given by triangles $\beta^+_{ij}=((i,j),(i+1,j),(i+1,j+1))$ and $\beta^-_{ij}=((i,j),(i,j+1),(i+1,j+1))$ with integer values $i,j\in\mathbb{Z}$. The space of edges $V^1$ contains the sequences $\alpha^1_{ij}=((i,j),(i+1,j))$, $\alpha^2_{ij}=((i,j),(i,j+1))$, $\alpha^{12}_{ij}=((i,j),(i+1,j+1))$. The space of vertices is given by $V=\ZZ^2$, and we denote $v_{ij}=(i,j)$.

For any fixed immersion $x\colon V\hookrightarrow X$ we may pull-back smooth bundles $\pi\colon Y\rightarrow X$ on $X$ to generate discrete bundles $\pi_x\colon Y_x\rightarrow V$ on $V$, where $Y_x=x^*Y=V\times_X Y$. In the case that $\pi\colon P\rightarrow X$ is a (smooth) principal $G$-bundle, also $\pi_x\colon P_x\rightarrow V$ is a discrete principal $G$-bundle. Consider the bundle of admissible rod motions $\pi\colon P\rightarrow \RR^2_{s,t}$. Fixing non-negative discretization parameters $\Delta s>0$, $\Delta t>0$, we have an immersion $x\colon (i,j)\mapsto (s_0+(i-j)\Delta s,t_0+(i+j)\Delta t)$, leading to the discrete principal bundle $\pi_x\colon P_x\rightarrow V$ on the set of vertices. For a discrete configuration $p_x\in\Gamma(P_x)$ the value $p_{x(i,j)}$ shall be an approximation to the configuration of slice $s_0+(i-j)\Delta s$ at time $t_0+(i+j)\Delta t$. Therefore configurations for $i+j=k\in\mathbb{Z}$ correspond to a discrete approximation of the rod configuration at fixed time $t_0+k\Delta t$, and configurations for $i-j=k\in\mathbb{Z}$ correspond to a discrete approximation of the time evolution of a fixed slice $s_0+k\Delta s$.

For each edge $\alpha\in V^1$ we consider $x\alpha\in X\times X$. Using the affine structure on $X$ we shall use $\Delta_X(x\alpha)\in T_{xv_0}X$ the usual forward difference operator $\Delta_X\colon X\times X\rightarrow TX$ determined by the affine structure $\Delta_X(x_0,x_1)=\overrightarrow{x_0x_1}\in T_{x_0}X$. For our specific choice of immersion $x\colon V\rightarrow \RR^2_{(s,t)}$ this becomes:
\begin{equation}\label{tangX}
\begin{aligned}
\Delta_X(x\alpha^1_{ij})&=(\Delta s)\cdot \left(\parcial{}{s}\right)_{x(i,j)} + (\Delta t)\cdot \left(\parcial{}{t}\right)_{x(i,j)}\\
\Delta_X(x\alpha^2_{ij})&=-(\Delta s)\cdot \left(\parcial{}{s}\right)_{x(i,j)} + (\Delta t)\cdot \left(\parcial{}{t}\right)_{x(i,j)}\\
\Delta_X(x\alpha^{12}_{ij})&=2 (\Delta t)\cdot \left(\parcial{}{t}\right)_{x(i,j)}
\end{aligned}
\end{equation}
in this sense edges $\alpha\in V^1$ can be associated with corresponding tangent vectors $\Delta_X(x\alpha)\in T_{xv_0}X$ where $v_0=\pi_0\alpha\in V$. Each edge $\alpha$ can be seen as a discrete approximation to the tangent vectors given in (\ref{tangX}).

Discrete $H$-reduced fields $(q_x,\psi_x)\in\Gamma(V,\Str{H}_x)\times \Gamma(V^1,\End^1P_x)$ shall be used as approximations to smooth $H$-reduced fields $(q(x),\chi(x))\in\Gamma(X,\Str{H})\times \Gamma(X,\con)$. Viewing $q_{x(i,j)}$ as approximation to $q(x(i,j))$ represents no problem, as both of them are on the same space $\Str{H}_{xv_{ij}}=\Str{H}_{x(i,j)}$. Using $\Gamma(V^1,\End^1 P_x)$ to approximate $\Gamma(X,\con)$ is more complex. It requires comparing elements in a nonlinear space $\End^1 P_{x\alpha}$ with elements in an affine subspace of $T^*X\otimes\at P$, a linearization mechanism that we introduce next:

\begin{prop}
	For any point $x\in X$ there exists a local diffeomorphism $\Delta_{\at}\colon \End P_x\rightarrow \at P_x$ from the bundle $(\bar\pi_0,\bar\pi_1)\colon \End P_x\rightarrow \{x\}\times X$ to the bundle $\overline{\dd\pi}\colon \at P_x\rightarrow T_xX$, projecting to $\Delta_X\colon \{x\}\times X\rightarrow T_xX$, defined on a neighborhood of $\Id_x\in\Gau P_x\subset\End P_x$, whose restriction to $\Gau P_x$ is the Lie group logarithm to its corresponding Lie algebra $\ad P_x$,
	\begin{equation}\label{variosdelta}
	\xymatrix{
		\Gau P_x \ar@{^{(}->}[r] \ar[d]^{\log} & \End P_x \ar[d]_{\Delta_{\at}}\ar[r]^{(\bar\pi_0,\bar\pi_1)} & \{x\}\times X \ar[d]^{\Delta_X} \\
		\ad P_x \ar[u]^{\exp}\ar@{^{(}->}[r] &\at P_x \ar[r]^{\overline{\dd\pi}} & T_xX
	}
	\end{equation}	
	and whose linear approximation at $\Id_x\in\Gau P_x\subset \End P_x=\bar\pi_0^{-1}(x)$ coincides with the natural identification $V^{\bar\pi_0}\End P\simeq \bar\pi_1^*\at P$ determined by the target trivialisation (\ref{tartri}).
\end{prop}
PROOF:\newline
Considering for the smooth principal bundle $P$ the identification $V^{\bar\pi_0}\End P\simeq \bar\pi_1^*\at P$, for the element $\Id_{x}\in\Gau_{x}\subset\End P_{x}$ we have $T_{\Id_{x}}\End^1 P_{x}\simeq \at P_{x}$, moreover, with this identification the projection $\dd\bar\pi_1\colon T_{\Id_{x}}\End P_{x}\rightarrow T_{x}X$ is identified with $\overline{\dd\pi}\colon \at_{x}\rightarrow T_{x}X$. Therefore $\ad P_{x}$ is identified with $\ker\dd\bar\pi_1=T_{\Id_{x}}\Gau P_{x}$. As $\ad P_{x}$ is the Lie algebra of the Lie group $\Gau P_{xv}$, there is an exponential mapping $\exp\colon \ad P_{x}\rightarrow \Gau P_{x}$, and the corresponding logarithm which has the property that the differential at $\Id_x$ is the identity mapping on $T_{\Id_x}\Gau P_x=\ad P_x$. We want to extend this situation to elements $\psi\in\End^1 P_x$ that do not project to $(x,x)$ (elements that don't belong to $\Gau P_x$).

The connection $\hat\chi_x\colon T_xX\rightarrow \langle\hat\sigma_x,\hat\tau_x\rangle\subset \at P_x$ together with $\Delta_X\colon \{x\}\times X\rightarrow T_xX$ determine a local diffeomorphism $\langle\hat\sigma_x,\hat\tau_x\rangle\simeq \{x\}\times X$, for which the projector $\overline{\dd\pi}\colon \langle\hat\sigma_x,\hat\tau_x\rangle\rightarrow T_xX$ gets identified with $\Delta_X\colon \{x\}\times X\rightarrow T_xX$ (because $\hat\chi$ is inverse to $\overline{\dd\pi}$ in the space $\langle\hat\sigma_x,\hat\tau_x\rangle$).

Consider now a 2-dimensional surface on $\End P_x$ passing through $\Id_x$ and whose tangent space at this point is precisely $\langle \hat\sigma_x,\hat\tau_x\rangle \subset \at P_x\simeq T_{\Id_x}\End P_x$. Using the implicit function theorem, such a surface determines a local section $\{x\}\times X\hookrightarrow \End P_x$ of the projector $(\bar\pi_0,\bar\pi_1)$ and  transforming $(x,x)$ into $\Id_x$. Composing with the previous local diffeomorphism $\langle\hat\sigma_x,\hat\tau_x\rangle\simeq \{x\}\times X$, we obtain a local immersion $\psi_x\colon \delta\in \langle\hat\sigma_x,\hat\tau_x\rangle\hookrightarrow \psi_x(\delta)\in\End P_x$ whose differential is identified with the inclusion $\langle\hat\sigma_x,\hat\tau_x\rangle\subset \at P_x$ and commuting with the corresponding projectors.

Combining the exponential mapping with our arbitrary choice of hypersurface, we get $(\delta,a)\in\langle\hat\sigma_x,\hat\tau_x\rangle\oplus \ad P_x=\at P_x\mapsto \psi_x(\delta)\circ\exp(a)\in\End^1P_x$. This is a smooth mapping, its restriction to the subspace $\langle\hat\sigma,\hat\tau\rangle$ is the immersed hypersurface chosen for $\End^1 P_x$, its restriction to the subspace $\ad P_x$ is the exponential mapping. Therefore it is a local diffeomorphism transforming $0\in\at P_x$ into $\Id_x\in\Gau P_x\subset\End P_x$, the associated tangent mapping at $\Id_x$ is the identity on $\at P_x$, and projects as $\Delta_X^{-1}\colon T_xX\rightarrow \{x\}\times X$. Taking a (locally defined) inverse of this mapping in a neighborhood of $\Id_x$ we obtain $\Delta_{\at}$ with all the conditions in our statement.$\square$

The expression above simplifies the question of how to compute a difference for two elements $\psi_0,\psi_1\in\End P_x$. If they share the source vertex $x$, we may not consider $\psi_1\circ\psi_0^{-1}$, that does not have $x$ as source. However, the linearization mechanism provided by $\Delta_{\at}$  would allow to compute $\Delta_{\at}(\psi_1)-\Delta_{\at}(\psi_0)\in\at P_x$ and to determine the difference as the only element $\psi\in\End P_x$ such that $\Delta_{\at}(\psi)=\Delta_{\at}(\psi_1)-\Delta_{\at}(\psi_0)$, which is well defined if both elements do not differ much from $\Id_x$.
\begin{define}\label{Def4.1}
	We call reduced forward difference operator at a point $x\in X$ any local diffeomorphism $\Delta_{\at}\colon \End P_x\rightarrow \at P_x$ with the characteristics given in the previous proposition. Locally, for elements that are near $\Id_x$ this determines $\psi\in\End P_x\mapsto \delta=\Delta_{\at}(\psi)\in \at P_x$ and $\delta\in \at P_x\mapsto \psi_\delta\in \End P_x$. 
\end{define}
The motivation for this definition can be found in \cite{CasiRodr17}. We may use such a local diffeomorphism $\Delta_{\at}\colon \End^1 P_{xv}\simeq \at_{xv}$ (or its inverse, that we call retraction mapping) to generate a section $\psi_x\in\Gamma(\End^1P_x)$, from any smooth connection $\chi=\dd t\otimes \tau+\dd s\otimes \sigma$
\begin{define}
	For any smooth connection $\chi\in\Gamma(\con)$, we call associated discrete connection $\psi_x\in\Gamma(V^1,\End^1P_x)$ the section determined by $\Delta_{\at}\left(\psi_{x\alpha}\right)=\langle \chi,\Delta_X(x\alpha)\rangle$ for each $\alpha\in V^1$
\end{define}
As each $\langle \chi,\Delta_X(x\alpha)\rangle$ projects by $\overline{\dd\pi}$ as $\Delta_X(x\alpha)$, using the property $\overline{\dd\pi}\circ\Delta_{\at}=\Delta_X\circ(\bar\pi_0,\bar\pi_1)$ we may state that the elements $\psi_{x\alpha}$ defined by this formula really project by $(\bar\pi_0,\bar\pi_1)$ into $x\alpha\in X\times X$, defining in this way a section $\psi_x\in\Gamma(V^1,\End^1 P_x)$.

Taking into account (\ref{tangX}) obtained for our specific immersion, we conclude:			
\begin{equation}\label{referencediscreteconnection}
\begin{aligned}
\Delta_{\at}\left(\psi_{x\alpha^1_{ij}}\right)&= (\Delta s)\cdot \sigma_{x(i,j)} + (\Delta t)\cdot \tau_{x(i,j)}\\
\Delta_{\at}\left(\psi_{x\alpha^2_{ij}}\right)&=-(\Delta s)\cdot \sigma_{x(i,j)} + (\Delta t)\cdot \tau_{x(i,j)}\\
\Delta_{\at}\left(\psi_{x\alpha^{12}_{ij}}\right)&= 2 (\Delta t)\cdot \tau_{x(i,j)}
\end{aligned}
\end{equation}
In the smooth variational formulation of hyperelastic rod dynamics there is a fixed principal connection $\hat\chi\in\Gamma(\con)$, characterized by $\langle \hat\chi_,\partial/\partial s\rangle=\hat\sigma$, $\langle \hat\chi,\partial/\partial t\rangle=\hat\tau$ ($\hat\sigma,\hat\tau$ the sections of the Atiyah bundle that represent the minimal kinetic and minimal strain energy). This induces a corresponding reference discrete connection $\hat\psi_x\in\Gamma(\End^1P_x)$ defined by:
\begin{equation*}
\begin{aligned}
\Delta_{\at}(\hat\psi_{x\alpha^1_{ij}})&= (\Delta s)\cdot \hat\sigma_{x(i,j)} + (\Delta t)\cdot \hat\tau_{x(i,j)}\\
\Delta_{\at}(\hat\psi_{x\alpha^2_{ij}})&=-(\Delta s)\cdot \hat\sigma_{x(i,j)} + (\Delta t)\cdot \hat\tau_{x(i,j)}\\
\Delta_{\at}(\hat\psi_{x\alpha^{12}_{ij}})&= 2 (\Delta t)\cdot \hat\tau_{x(i,j)}
\end{aligned}
\end{equation*}
Using this reference discrete connection, the remaining ones have a simpler description:
\begin{define}
	For any reference discrete connection $\hat\psi\in\Gamma(\End^1P_x)$ we call linearization of any other discrete connection $\psi\in\Gamma(\End^1P_x)$ with respect to $\hat\psi$ the section $\delta\in\Gamma(V^1,\pi_0^*\ad P_x)$ determined by $\delta_{x\alpha}=\Delta_{\at}(\psi_{x\alpha})-\Delta_{\at}(\hat\psi_{x\alpha})$
\end{define}
In our particular case
\begin{equation}\label{definedeltas}
\begin{aligned}
\delta_{x\alpha^1_{ij}}&=\Delta_{\at}(\psi_{x\alpha^1_{ij}})- (\Delta s)\cdot \hat\sigma_{x(i,j)} - (\Delta t)\cdot \hat\tau_{x(i,j)}\\
\delta_{x\alpha^2_{ij}}&=\Delta_{\at}(\psi_{x\alpha^2_{ij}})+(\Delta s)\cdot \hat\sigma_{x(i,j)} - (\Delta t)\cdot \hat\tau_{x(i,j)}\\
\delta_{x\alpha^{12}_{ij}}&=\Delta_{\at}(\psi_{x\alpha^{12}_{ij}})- 2 (\Delta t)\cdot \hat\tau_{x(i,j)}
\end{aligned}
\end{equation} 	
With these expressions, the values $\hat\delta_{x\alpha}$ associated to the reference discrete connection $\hat\psi$ described in (\ref{referencediscreteconnection}) are clearly  $\delta_{x\alpha}=0$ at each edge.

As $\psi_{x\alpha}$ projects into $x\alpha$, in (\ref{definedeltas}) we are substracting elements that project into $\Delta_X(x\alpha)$, this leads to a section $\delta\in\Gamma(V^1,\pi_0^*\ad P_x)$. Conversely, any such section, determines $\Delta_{\at}(\psi_{x\alpha})$ for each $\alpha\in V^1$ and using the retraction $\Delta_{\at}^{-1}$, determines a section $\psi\in\Gamma(\End^1 P_x)$. The $H$-reduced discrete Lagrangian for a hyperelastic rod is given by (\ref{disredlagrod}). We may observe also from (\ref{definedeltas}) and (\ref{referencediscreteconnection}) that:
\begin{equation*}
\begin{aligned}
\sigma_{x(i,j)}-\hat\sigma_{x(i,j)}&=\frac{\delta_{x\alpha^{12}_{ij}}-2\delta_{x\alpha^{2}_{ij}}}{2\Delta s}=\frac{-\delta_{x\alpha^{12}_{ij}}+2\delta_{x\alpha^{1}_{ij}}}{2\Delta s} \\
\tau_{x(i,j)}-\hat\tau_{x(i,j)}&=\frac{\delta_{x\alpha^{12}_{ij}}}{2\Delta t} 
\end{aligned}
\end{equation*} 
For any facet $\beta=(v_0,v_1,v_2)\in V^2$ the edge $(v_0,v_1)\in V^1$ is one of the form $\alpha^1_{ij}$ or $\alpha^2_{ij}$, and the edge $(v_0,v_2)$ is one of the form $\alpha^{12}_{ij}$. In any of the two cases the formulas above allow to determine the difference $\tau-\hat\tau$ or $\sigma-\hat\sigma$ (used in the smooth $H$-reduced Lagrangian $\bar\ell$), at all the points $x(i,j)$. Therefore considering (\ref{disredlagrod}) we arrive to the following:
\begin{define}\label{discretisedL}
	We call linearized $H$-reduced discrete rod motion any pair pair of discrete fields $(q_x,\delta_x)\in\Gamma(\Str{H}_x)\times \Gamma(V^1,\pi_0^*\ad P_x)$. For any $\beta\in V^2$ we call hyperelastic Lagrangian for linearized $H$-reduced rod motions, the following:
	\begin{equation*}
	\begin{aligned}
	\bar\ell_{x\beta}&(q_{xv_0},\delta_{xv_0v_1},\delta_{xv_0v_2})=\frac12 \left\| \frac{\delta_{xv_0v_2}}{2\Delta t} \right\|_K^2\Delta s\Delta t- \frac12 \left\| \frac{\delta_{xv_0v_2}-2\delta_{xv_0v_1}}{2\Delta s} \right\|^2_W\Delta s\Delta t+ \omega(q_{xv_0})\Delta s\Delta t=\\
	&=\frac{1}{4\Delta s\Delta t}\left( \frac12 (\Delta s)^2\left\| \delta_{xv_0v_2} \right\|_K^2- \frac12 (\Delta t)^2\left\| \delta_{xv_0v_2}-2\delta_{xv_0v_1} \right\|^2_W+ 4 \omega(q_{xv_0})(\Delta s)^2(\Delta t)^2\right)
	\end{aligned}
	\end{equation*}
	This hyperelastic Lagrangian determines a discrete action functional, for any choice of a finite domain $K\subset V^2$:
	\begin{equation*}
	\LL_K(q_x,\delta_x)=\sum_{\beta\in K}\bar\ell_{x\beta}(q_{x\beta},\delta_{x\beta})
	\end{equation*}
	where $q_{x\beta}$, $\delta_{x\beta}$ are the obvious extensions to any facet $\beta\in X_2$ of $q_x\in\Gamma(V,\Str{H}_x)$, $\delta_x\in\Gamma(V^1,\pi_0^*\ad P_x)$
\end{define}
We observe that the identification of linearized objects with nonlinearized $H$-reduced objects determined by $\Delta_{\at}$ and its local inverse shows that this discrete Lagrangian determines an $H$-reduced discrete Lagrangian on the bundle $RJP_x$. This $H$-reduced discrete Lagrangian on $RJP_x$ also determines an $H$-invariant discrete Lagrangian on the discrete jet bundle $JP_x$. In this case the action functional determines a variational principle, and critical sections $p\in\Gamma(P_x)$ are characterized by discrete Euler-Lagrange equations (\ref{discreteEL}). Moreover, solutions of these equations satisfy conservation laws as described in Theorem \ref{Noether}. To impose these equations for a (non-reduced) discrete field is equivalent to impose that the corresponding $H$-reduced discrete field $(q_x,\psi_x)$ satisfies the corresponding discrete Euler-Poincaré equations (\ref{dEP}), together with the conditions of discrete flatness (\ref{dflat}) and parallelism (\ref{dpara}). 
\begin{define}
	We say a linearized $H$-reduced discrete rod motion any pair of sections $(q_x,\delta_x)\in\Gamma(\Str{H}_x)\times \Gamma(V^1,\pi_0^*\ad P_x)$ is admissible if its associated $H$-reduced discrete field $(q_x,\psi_x)$ satisfies the corresponding conditions of discrete flatness (\ref{dflat}) and parallelism (\ref{dpara}).
\end{define}

Conditions (\ref{dflat}) and (\ref{dpara}) are simple to express in terms of discrete $H$-reduced fields $(q,\psi)$. On the other hand the specific expression of the discrete Lagrangian in terms of $(q_x,\psi_x)\in\Gamma(\Str{H}_x)\times \Gamma(\End^1 P_x)$ or in terms of $p_x\in\Gamma(P_x)$ is complicated and strongly depends on the particular choice of $\Delta_{\at}$. Moreover, for these fields the specific form of discrete Euler-Lagrange equations (\ref{discreteEL}) or Euler-Poincaré equations (\ref{dEP}) imply the addition of several partial derivatives of the discrete Lagrangian, which makes the local coordinate expressions almost impossible to explore. 

Working in linearized coordinates the integration of the Euler-Poincaré equations becomes much simpler. First of all, suppose that all components $q_{xv}$, $\delta_{x\alpha}$ of the linearized $H$-reduced discrete field $(q_x,\delta_x)\in\Gamma(\Str{H}_x)\times\Gamma(V^1,\pi_0^*\ad P_x)$ are known at vertices and edges contained in the semi-space $i+j\leq k$. For any new vertex $v_2=(i_1,j_1)$ with $i_1+j_1=k+1$,  Euler-Poincaré equation associated to $\bar\ell_{x\beta}$ at $v_0=(i_1-1,j_1-1)=(i_0,j_0)$ depends on several $H$-structure and linearized connection parameters, all of them known, except for a single one $\delta_{xv_0v_2}$. In fact in its trivialized ($\ad^*P_x$-valued) form Euler-Poincaré equation has a component in $\ad^* P_{xv_0}$ that reads as:
\begin{equation}\label{EPeq}
2(\Delta s)^2K(\delta_{xv_0v_2})-(\Delta t)^2W(\delta_{xv_0v_2}-2\delta_{xv_0v_1})-(\Delta t)^2W(\delta_{xv_0v_2}-2\delta_{xv_0\bar v_1})=\text{known terms}
\end{equation}
where $v_1=v_0+(1,0)$, $\bar v_1=v_0+(0,1)$, $v_2=v_0+(1,1)$, and known terms refer to elements that depend on $\delta_{x\alpha}$ and $q_{xv}$ for vertices and edges $v,\alpha$ contained in the semi-space $i+j\leq k$. The specific expression involves, however the transformation with $\ad^*_{\psi}$ of several elements of the form $K(\delta),W(\delta)\in\ad^*P$, and its addition.

\begin{remark}\label{remark02}
	As indicated in Remark \ref{remark01} discretization methods of smooth lagrangians in mechanics rely on the choice of a quadrature rule and a certain numerical scheme for ordinary differential equations. In geometric mechanics on Lie groups lagrangian densities can be trivialised to the form $\ell(t,g,\chi)$ with $\chi=g^{-1}\dot g\in \Lie G$. In the discretization procedure one uses the exponential mapping or any other retraction mapping $\tau\colon \Lie G\rightarrow G$ (with $\dd_0\tau=\Id_{\Lie G}$) to generate a Lie-algebra element $\chi_0=\frac{\tau^{-1}(g_0^{-1}g_1)}{t_1-t_0}$ for any pair of configurations $(t_0,g_0), (t_1,g_1)$ on $\RR\times G$. This element $\chi_0\in\Lie G$ is considered a good approximation to $g^{-1}\dot g(t_0)$ if $g_0,g_1$ are considered good approximations to $g(t_0)$, $g(t_1)$ for some smooth trajectory $g(t)$, because for any smooth trajectory $g(t)$ and using $g_1=g(t_1)$, $g_0=g(t_0)$ in the limit $t_1\to t_0$, the Lie algebra element $\chi_0(t_0,g(t_0),t_1,g(t_1))$ converges to the element $g^{-1}\dot g(t_0)$ determined at $t_0$ by the smooth trajectory $\chi(t)=g^{-1}\dot g(t)$.
	
	Therefore $\ell(t_0,g_0,\chi_0(t_0,g_0,t_1,g_1))\cdot (t_1-t_0)$ is considered a good approximation to $\int_{t_0}^{t_1} \ell(t,g(t),\chi(t))\dd t$. To compute the error of this approximation, we only need to observe that we applied the rectangle quadrature rule $\int_{t_0}^{t_1}f(t)\dd t\simeq f(t_0)\cdot (t_1-t_0)$ at $t_0$ for the integration in the interval $[t_0,t_1]$ to obtain $\ell(t_0,g(t_0),\chi(t_0))(t_1-t_0)\simeq \int_{t_0}^{t_1} \ell(t,g(t),\chi(t))\dd t$, and the substitution of approximate values $g(t_0)\simeq g_0$, $g^{-1}\dot g(t_0)\simeq \chi_0$. In the case described above the choice of retraction mapping $\tau$ can be seen as a one-step numerical scheme that allows to recover $g_1$ from initial conditions $(g_0,\chi_0)$, solving for time $t_1$ the ordinary differential equation $g^{-1}\dot g=\chi$ with initial conditions $g(t_0)=g_0$, $g^{-1}\dot g(t_0)=\chi_0$. The choice of the discrete Lagrangian $\ell(t_0,g_0,\chi_0)\cdot (t_1-t_0)$ arises then from a choice of a rectangle quadrature rule to compute integrals on $[t_0,t_1]$ and a numerical scheme (determined by the choice of a retraction mapping $\tau$) to solve differential equation $g^{-1}\dot g=\chi$. This simple scheme has been used as foundation and extended to include more complex quadrature rules with several nodes \cite{LeoShi12} or higher order lagrangians. This general structure of discretization allows to translate existing
	theoretical results and techniques in approximation theory, and simplifies the numerical analysis of time-integration methods for a large family of variational integrators.
	
	Transferring these ideas to discrete (reduced) field theories, our method to discretize the Lagrangian density of a field theory (definition \ref{discretisedL}) is based on \cite{CasiRodr17} (see Definition 5.12 and 8.3 in this reference). It relies on a generalization for principal bundles of the notion of retraction mapping for Lie groups. This generalization is the notion of reduced forward difference operator $\Delta_{\at}$ given in definition \ref{Def4.1}, which fixes, for any facet $\beta\in V^n$, a one-to-one correspondence (see \cite{CasiRodr17}) between elements in the reduced jet space $RJP_{x\beta}$, and  elements $(q_{x_0},\chi_{x_0})\in \Str{H}_{x_0}\times \con_{x_0}$ for $x_0=\pi_0(x\beta)$. The second main ingredient is a very simple cubature rule for a simplicial domain, namely the action functional $\int_{[x\beta]} \ell(x,q(x),\chi(x))\vol_X$ on a simplicial domain $[x\beta]\subset X$ with vertices $x\beta\in X^n$ is approximated taking the value of the lagrangian $\ell(x_0,q(x_0),\chi(x_0))$ at a given point $x_0=\pi_0( x\beta)=xv_0\in X$, multiplied with the volume of the simplicial domain $\int_{[x\beta]}\vol_X$. These approximations generate the smooth lagrangian density when one considers the limit case with $\Delta s,\Delta t$ tending to $0$. However this cubature rule is not essential, and the hyperelastic Lagrangian for linearized $H$-reduced rod motions described in definition \ref{discretisedL} using a cubature rule on triangles, with a single node, could be changed by other alternative cubature rules on simplicial domains, that may have higher order of precision. In this regard, due to our simplicial choice of discrete space, we may consider different Newton-Cotes cubature rules on principal lattices, which are described by Nicolaides \cite{Nicolaides72}, and much studied in the literature (see references in \cite{GascaSauer}), with known precision order.

\end{remark}

The important aspect now is that Euler-Poincaré equations (\ref{EPeq}) implicitly determine $\psi_{xv_0v_2}$, normally in a non-linear way. However working with the linearized field $\delta$, assuming that $(\Delta s)^2K-(\Delta t)^2W$ is non-degenerate, the element $\delta_{xv_0v_2}$ is obtained solving a system of linear equations (we get a particular simple case of implicit equations). Taking into account that $K$ is positive-definite, this can always be solved in a unique way, as long as the ratio $\Delta s/\Delta t$ does not lie in a finite set of critical values that would make $(\Delta s)^2K-(\Delta t)^2W$ degenerate.

Having obtained $\delta_{xv_0v_2}$, equations (\ref{definedeltas}) and the inverse morphism $\Delta_{\at}^{-1}$ can be used to generate $\psi_{xv_0v_2}$. The parallelism condition  $q_{xv0}\psi_{xv_0v_2}=q_{xv_2}$ (\ref{dpara}) determines the $H$-structure component at $xv_2$. The flatness condition $\psi_{xv_0v_1}\psi_{xv_1v_2}=\psi_{xv_0v_2}$ (\ref{dflat}) determines then the discrete connection component $\psi_{xv_1v_2}$ which, using $\Delta_{\at}$, allows to compute $\delta_{xv_1v_2}$.

The application of the previous steps allows to recover all discrete $H$-structure and connection components at vertices and edges on the semi-space $i+j\leq k+1$, whenever we know these components on the semi-space $i+j\leq k$. We may observe that the transition from $i+j\leq k$ to $i+j\leq k+1$ corresponds to the recovery or a discrete rod configuration at time $t+\delta t$ if we know the configurations in time $t$ and in previous times.

\section{Conclusion}

In the different sections we presented the geometrical formulation of hyperelastic rod theory. This formulation led us to the study of variational principles in principal $G$-bundles, how these fields can be trivialized using the bundle of principal connections, and for the case of Lagrangians that admit a subgroup $H\subseteq G$ of symmetries, we showed that the corresponding reduced fields can be used to describe the dynamics of the rod. In this case the relevant equations are not the vanishing of Euler-Lagrange tensor, but the vanishing of Euler-Poincaré tensor, together with certain compatibility conditions (parallelism of the $H$-structure and flatness of the connection).

With this background, we explored the possibility to give a formulation of all the needed geometrical tools in a discrete setting. The first relevant choice is the consideration of abstract simplicial complexes as discrete base manifold. In this case most of the geometrical ideas needed for a variational theory are determined from this topological choice, and no additional tool seems necessary. The whole variational theory for the reduction of variational principles in principal $G$-bundles can be formulated then in a discrete setting.

Finally, we found two tools needed to discretise a smooth variational principle to generate a discrete one. Firstly, a simple immersion $x$ of vertices from a simplicial complex into the base manifold of the smooth theory. Secondly the reduced forward difference operator $\Delta_{\at}$, that allows to linearize elements of the Ehresmann groupoid as elements of the Atiyah vector bundle. We show that this tool can be used to formulate discrete variational principles related to the smooth variational principle of the hyperelastic rod dynamics. We also show that with our particular choice of discrete space and discretized Lagrangian, the problem of computing solutions of the discrete Euler-Poincaré equation can be performed, obtaining discrete solutions that have corresponding Noether current conservation properties. With non-geometrical tools, Euler-Poincaré equations could be discretized, becoming nonlinear implicit equations that could be solved in approximate way using some iterative method. However small errors in the solution of the implicit equations, with these techniques do not warrant that some associated Noether current is conserved. Even though the conceptual machinery was considerable, the computational cost for solving the final equations is really low: There appear no implicit equations except for a single system of linear equations, always with the same coefficients $(\Delta s)^2K-(\Delta t)^2W$. The main object $\Delta_{\at}$ and its inverse may be chosen almost arbitrarily, reducing in this way its computational cost.

From this point, two main tasks arise: In a particular case, like this discrete hyperelastic rod model, how do these techniques behave from a numerical point of view with respect to other techniques that have a similar computational cost, in this kind of problem? Such a study is amenable to techniques similar to those performed in geometric mechanics on Lie groups, reducing the analysis to convenient choices of a cubature rule and a ``retraction mapping'' (here a reduced forward difference operator). As indicated in remark \ref{remark02} the foundations are laid for the introduction of new cubature rules (with more than one node) on simplicial domains (see for example \cite{GascaSauer}). In the same manner as retraction mappings are used to approximate frame-independent representation of velocities on Lie groups, a choice of reduced forward difference operator $\Delta_{At}$ can be used to approximate jets or reduced jets (principal connections) in field theories on principal bundles. The combined analysis of precision degree for the cubature rule on a simplex and of precision degree for the reduced forward difference operator should determine error bounds for the smooth action functional on a critical field, on a simplicial domain, with respect to the discrete action functional on the values of this section at the vertices of that simplicial domain. These error bounds would then lead to error bounds when we compare the exact discrete lagrangian, with the discretised lagrangian obtained with an specific cubature rule and $\Delta_{At}$. 

A second question is, for the general case of variational principles in principal bundles, can we introduce alternative elements in the theory that lead to a more faithful discrete representation of the smooth variational principle? That is, how do our choices of discrete space (simplicial complex), cubature rule, and reduced forward difference operator influence corresponding error bounds for euler-lagrange equations, conserved quantities, and critical fields for the corresponding discrete and smooth formalisms?

\section{Acknowledgments}
Fundação para a Ciência e a Tecnologia (Portuguese Foundation for
Science and Technology) supported Ana Casimiro through the project UID/MAT/00297/2013 (Centro de Matemática e Aplicações), also supporting  César Rodrigo through the project UID/MAT/04561/2013 (Centro de Matemática, Aplicações Fundamentais e Investigação Operacional of
Universidade de Lisboa CMAF-CIO).



\begin{thebibliography}{99}
	\bibitem{AMS08} P.A.~Absil, R.~Mahony, and R.~Sepulchre. \emph{Optimization algorithms on matrix manifolds}.
	Princeton University Press (2008) 240pp
	
	\bibitem{Adler} R.L.~Adler, J.P.~Dedieu, J.Y.~Margulies, M.~Martens, M.~Shub. \emph{Newton's method on Riemannian manifolds and a geometric model for the human spine}. IMA J. Numer. Anal. 22(3), (2002) 359--390.
	
	\bibitem{Antman94}  S.S.~Antman, \emph{Nonlinear problems of elasticity}. In: Applied Mathematical Sciences 107, Springer Verlag	(1994)	
	
	\bibitem{Atiyah} M.~F.~Atiyah. \emph{Complex analytic connections in fibre bundles}. Trans. Amer. Math. Soc. 85, (1957)  181–-207
	
	\bibitem{BobeSuri99} A.I.~Bobenko, Yu.B.~Suris. \emph{Discrete Lagrangian reduction, discrete Euler–Poincaré equations, and semidirect products}. Lett. Math. Phys. 49.1 (1999) 79--93.
	
	\bibitem{CannWein99} A.~Cannas~da~Silva, A.~Weinstein. \emph{Geometric models for noncommutative algebras}. In: Berkeley Mathematics Lecture Notes
	Volume: 10 (1999) 184pp.
	
	
	\bibitem{Capr14} S.~Capriotti, \emph{Differential geometry, Palatini gravity and reduction}. J. Math. Phys. 55 (2014) 012902	
	
	\bibitem{CasiRodr12} A.C.~Casimiro, C.~Rodrigo, \emph{First variation formula and conservation laws in several independent discrete variables.} J.~Geom.~Phys.~62.1 (2012), 61--86.
	
	\bibitem{CasiRodr12b} A.C.~Casimiro, C.~Rodrigo, \emph{First variation formula for discrete variational problems in two independent variables.} RACSAM~Rev.~R.~Acad.~A~106(1) (2012), 111--135.
	
	
	\bibitem{CasiRodr17} A.C.~Casimiro, C.~Rodrigo, \emph{Reduction of Forward difference operators in principal G-bundles}. Stat. Optim. Inf. Comput. 6, March 2018, pp 42–85
	
	\bibitem{CasiRodr17b} A.C.~Casimiro, C.~Rodrigo, \emph{Variational integrators for reduced field equations}. Stat. Optim. Inf. Comput. 6, March 2018, pp 86–115
	
	\bibitem{Castrillon12} M.~Castrillón~López, \emph{Field theories: reduction, constraints and variational integrators.} RACSAM~Rev.~R.~Acad.~A 106(1) (2012), 67--74
	
	\bibitem{CasChaGar13} M.~Castrill\'on~L\'opez, P.~Chac\'on, P.L.~Garc\'{\i}a, \emph{Lagrange-Poincar\'e reduction in affine principal bundles}, J. Geom. Mech. 5 (vol 4), 399--414
	
	\bibitem{CasGarRat01} M.~Castrill\'{o}n~L\'{o}pez, P.L.~Garc\'{\i}a, T.S.~Ratiu, \emph{Euler-Poincar\'{e} reduction on principal bundles}, Lett. Math. Phys. 58 (2001) 167--180
	
	\bibitem{CasGarRod13} M.~Castrill\'{o}n~L\'{o}pez, P.L.~Garc\'{\i}a, C.~Rodrigo, \emph{Euler-Poincar\'{e} reduction in principal bundles by a subgroup of the structure group}, J. Geom. Phys. 74 (2013) 352--369
	
	\bibitem{ChriMuntOwre11} S.~H.~Christiansen, H.~Z.~Munthe-Kaas, B.~Owren, \emph{Topics in structure-preserving discretization.}  Acta Numer. 20 (2011), 1--119.
	
	\bibitem{CortMart01} J.~Cort\'{e}s, S.~Mart\'{\i}nez, \emph{Non-holonomic integrators}. Nonlinearity 14 (2001) 1365--1392.
	
	\bibitem{DemGayLeyObeRatWei} F.~Demoures, F.~Gay-Balmaz, S.~Leyendecker, S.~Ober-Blöbaum, T.S.~Ratiu, Y.~Weinand. \emph{Discrete variational Lie group formulation of geometrically exact beam dynamics}. Numer. Math. 130(1), (2015) 73--123.
	
	\bibitem{DemoGayRat} F.~Demoures, F.~Gay-Balmaz, T.S.~Ratiu. \emph{Multisymplectic variational integrators and space/time symplecticity}. Anal. Appl. 14(03), (2016) 341--391.
	
	\bibitem{Ehresmann50} C.~Ehresmann, \emph{Les connexions infinitésimales dans un espace fibré différentiable}, Coll. de
	Topologie, Bruxelles, CBRM (1950), 29–-55.
	
	\bibitem{ElliGayHolmRati11} D.C.P.~Ellis, F.~Gay-Balmaz, D.D.~Holm, T.S.~Ratiu, \emph{Lagrange-Poincar\'{e} field equations}, J. Geom. Phys. 61(11) (2011) 2120-2146.
	
	\bibitem{FernGarcRodr04} A.~Fern\'{a}ndez, P.L.~Garc\'{\i}a, C.~Rodrigo. \emph{Stress–energy–momentum tensors for natural constrained variational problems.} J. Geom. Phys. 49(1) (2004) 1--20.
	
	\bibitem{FerGarRod12} A.~Fern\'{a}ndez, P.L.~Garc\'{\i}a, C.~Rodrigo, \emph{Variational integrators in discrete vakonomic mechanics}. RACSAM~Rev.~R.~Acad.~A~106.1 (2012), 137--159.
	
	\bibitem{FernZucc13} J.~Fern\'{a}ndez, M.~Zuccalli. \emph{A geometric approach to discrete connections on principal bundles.} J. Geom. Mech. 5(4) (2013) 433--444.
	
	\bibitem{GascaSauer} M.~Gasca,T.~Sauer. \emph{Polynomial interpolation in several variables} Adv. Comput. Math. 12 (2000) 377. 
	
	\bibitem{GawlMullPavlMarsDesb11} E.S.~Gawlik, P.~Mullen, D.~Pavlov, J.E.~Marsden, M.~Desbrun. \emph{Geometric, variational discretization of continuum theories}. Physica D 240(21), (2011) 1724--1760.
	
	\bibitem{GayRatTro} F.~Gay-Balmaz, T.S.~Ratiu, C.~Tronci, \emph{Equivalent Theories of Liquid Crystal Dynamics}, Arch. Ration. Mech. Anal. 210(3), 773-811 (2013)	
	
	\bibitem{GoldStern73} H.~Goldschmidt, Sh.~Sternberg, Differential geometry, \emph{The Hamilton-Cartan formalism in the Calculus of Variations}, Ann. Inst. Four. 23(1) 203--267 (1973)
	
	\bibitem{GuoWu03} H.-Y.~Guo, K.~Wu, \emph{On variations in discrete mechanics and field theory}, J.~Math.~Phys 44(12) (2003) 5978--6004.
	
	\bibitem{HaiLubWan06} E.~Hairer, C.~Lubich, G.~Wanner, \emph{Geometric numerical Integration: Structure-Preserving Algorithms for Ordinary Differential Equations.} Springer Series in Computational Mathematics 31. Springer Verlag Berlin Heidelberg, New York, 2004.
	
	\bibitem{IseMunNorZan05} A.~Iserles, H.Z.~Munthe-Kaas, S.~P.~Nørsett, A.~Zanna. \emph{Lie-group methods}. Acta Numer. 9 (2000) 215--365.
	
	\bibitem{Kobilarov14} M.~Kobilarov. \emph{Solvability of Geometric Integrators for Multi-body Systems}. Multibody Dynamics. Springer International Publishing, 2014. 145--174.
	
	\bibitem{KoMa10} M.~B.~Kobilarov, J.~E.~Marsden. \emph{Discrete geometric optimal control on Lie groups} IEEE Transactions on Robotics 27.4 (2011) 641-655.
	
	\bibitem{Leok05} M.~Leok, J.E.~Marsden, and A.D.~Weinstein. \emph{A discrete theory of connections on principal bundles.} arXiv preprint math/0508338 (2005).
	
	\bibitem{LeoShi12} M.~Leok, T.~Shingel, \emph{General Techniques for Constructing Variational Integrators} Front.~Math.~China~7(2) (2012) 273–-303.
	
	\bibitem{LeonDiegSant04} M.~de~Le\'{o}n, D.~Mart\'{\i}n~de~Diego, A.~Santamar\'{\i}a~Merino, \emph{Geometric integrators and nonholonomic mechanics}. J.~Math.~Phys. 45(3) (2004) 1042--1064.
	
	\bibitem{LewMarsOrtiWest03} A.~Lew, J.E.~Marsden, M.~Ortiz, M.~West, \emph{Asynchronous variational integrators.} Arch.~Rat.~Mech.~Anal.~167 (2003), 85--146.
	
	\bibitem{MarsPekaShko00} J.E.~Marsden, S.~Pekarsky, S.~Shkoller. \emph{Symmetry reduction of discrete Lagrangian mechanics on Lie groups}. J.~Geom.~Phys. 36.1 (2000) 140--151.
	
	\bibitem{MarsWest01} J.~E.~Marsden, M.~West, \emph{Discrete mechanics and variational integrators}. Acta~Numer. 10 (2001) 357--514.
	
	\bibitem{McLa} R.~McLachlan, M.~Perlmutter. \emph{Integrators for nonholonomic mechanical systems} J. Nonlinear Sci. 16.4 (2006) 283--328.
	
	\bibitem{McLaQuis06} R.I.~Mc~Lachlan, G.R.W.~Quispel, \emph{Geometric integrators for ODEs.} J.~Phys.~A:~Math.~Gen.~39 (2006), 5251--5285.
	
	\bibitem{Nicolaides72} R.~A.~Nicolaides. \emph{On a class of finite elements generated by Lagrange interpolation}, SIAM J. Numer. Anal. 9 (1972), 435–445.
	
	\bibitem{Saunders} D.J.~Saunders. \emph{The geometry of jet bundles}. London Mathematical Society Lecture Note Series 142. Cambridge University Press, (1989)
	
	\bibitem{Vanker07} J.~Vankerschaver. \emph{Euler-Poincaré reduction for discrete field theories}. J.~Math.~Phys. 48(3), (2007) 032902.
	
	\bibitem{VankCantr07} J.~Vankerschaver, F.~Cantrijn. \emph{Discrete Lagrangian field theories on Lie groupoids.} J.~Geom.~Phys. 57(2) (2007) 665--689.
	
	\bibitem{VanLiaLeo12} Vankerschaver, C.~Liao, M.~Leok. \emph{Generating functionals and Lagrangian partial differential equations} J.~Math.~Phys. 54 (2013)  082901.
	
	\bibitem{WendMars97} J.M.~Wendlandt, J.E.~Marsden, \emph{Mechanical integrators derived from a discrete variational principle.} Physica~D~106 (1997), 223--246.
	
\end{thebibliography}
\end{document}